%% file: main.tex
\documentclass[twocolumn,nolinenumbers]{aastex62}
\usepackage[utf8]{inputenc}
\usepackage{xspace}
\usepackage{amsmath}
\usepackage{url}
\usepackage{graphicx}
\usepackage{float}
\usepackage{lineno}
\nolinenumbers

\def\gy#1{\textcolor{olive}{GY: #1}}

\shorttitle{The Galactic population of magnetars} 
\shortauthors{Sautron, McEwen et al.}

\begin{document}
\nolinenumbers

\title{The Galactic population of magnetars : a simulation-based inference study}

\authorcomment1{The two first authors contributed equally.}
\author[0009-0006-7913-1186]{M.~Sautron}
\affiliation{Université de Strasbourg, CNRS, Observatoire astronomique de Strasbourg, UMR 7550, 67000 Strasbourg, France}

\author[0000-0001-5481-7559]{A.~E.~McEwen}
\affiliation{University of Maryland Baltimore County, Baltimore, MD 21250, USA}

\author[0000-0002-7991-028X]{G.~Younes}
\affiliation{Astrophysics Science Division, NASA/GSFC, Greenbelt, MD 20771, USA}
\affiliation{Department of Physics, The George Washington University, Washington, DC 20052, USA}

\author[0000-0003-3790-8066]{J.~Pétri}
\affiliation{Université de Strasbourg, CNRS, Observatoire astronomique de Strasbourg, UMR 7550, 67000 Strasbourg, France}

\author[0000-0001-7833-1043]{P.~Beniamini}
\affiliation{Department of Natural Sciences, The Open University of Israel, P.O. Box 808, Ra’anana 4353701, Israel}
\affiliation{Astrophysics Research Center of the Open University (ARCO), The Open University of Israel, P.O. Box 808, Ra’anana 4353701, Israel}
\affiliation{Department of Physics, The George Washington University, Washington, DC 20052, USA}

\author[0000-0002-1169-7486]{D.~Huppenkothen}
\affiliation{Anton Pannekoek Institute for Astronomy, University of Amsterdam, Science Park 904, 1098XH Amsterdam, The Netherlands}

\correspondingauthor{M.~Sautron}
\email{matteo.sautron@gmail.com}
\correspondingauthor{A.~E.~McEwen}
\email{almcewen0@hotmail.com}

\begin{abstract}
Population synthesis modeling of the observed dynamical and physical properties of a population is a highly effective method for constraining the underlying birth parameters and evolutionary tracks. In this work, we apply a population synthesis model to the canonical magnetar population to gain insight into the parent population. We utilize simulation-based inference to reproduce the observed magnetar population with a model which takes into account the secular evolution of the force-free magnetosphere and magnetic field decay simultaneously and self-consistently. Our observational constraints are such that no magnetar is detected through their persistent emission when convolving the simulated populations with the XMM-Newton EPIC-pn Galactic plane observations, and that all of the $\sim$30 known magnetars are discovered through their bursting activity in the last $\sim50$~years. Under these constraints, we find that, within 95\% credible intervals, the birth rate of magnetars to be $1.8^{+2.6}_{-0.6}$~kyr$^{-1}$, and lead to having $10.7^{+18.8}_{-4.4}$~\% of neutron stars born as magnetars. We also find a mean magnetic field at birth ($\mu_b$ is in T) $\log\left(\mu_b\right) = 10.2^{+0.1}_{-0.2}$, a magnetic field decay slope $\alpha_d = 1.9 ^{+0.9}_{-1.3}$, and timescale $\tau_d = 17.9^{+24.1}_{-14.5}$~kyr, in broad agreement with previous estimates. We conclude this study by exploring detection prospects: an all-sky survey with XMM-Newton would potentially allow to get around 7 periodic detections of magnetars, with approximately 150 magnetars exceeding XMM-Newton's flux threshold, and the upcoming AXIS experiment should allow to double these detections.
\end{abstract}

\section{Introduction}

Magnetars are a peculiar subgroup of neutron stars, with magnetic fields much stronger than typical neutron stars ($10^{9} - 10^{11}$~T, while normal pulsars have magnetic fields closer to $10^{8}$~T). Given that such B-fields are unattainable in terrestrial laboratories, magnetars are the only means for studying the effects of ultra-high magnetic fields on matter, radiation, and fundamental physical processes. Naturally, questions about the formation of such neutron stars emerge. One explanation utilizes a dynamo process during the birth of the neutron star which dramatically increases the magnetic field after core collapse \citep{dt+92,bgr+24}. Another explanation (known as the fossil field hypothesis) suggests that the magnetic field of the progenitor star was already large and subsequently amplified during the core-collapse supernova event by magnetic flux conservation.

Known magnetars are typically young objects, evidenced by their rapid rotational decay compared to normal pulsars, and the placement of some within X-ray bright, young supernova remnants. They have long spin periods ($P \approx 1-15\,{\rm s}$) and large spin period derivatives ($\dot{P} \approx 10^{-15}-10^{-11}\,{\rm s/s}$). Under the assumption of a rotating-dipole in vaccum, these rotational characteristics indeed translate to dipole fields of the order of $10^{11}$~T measured at the equator. Most known magnetars are bright X-ray emitters, with X-ray luminosities in the range of $10^{25}-10^{28}$~W, which at times exceed the corresponding rotational energy losses $\dot{E}$ by a few orders of magnitude. Hence, the emission process from these objects has been demonstrated to be driven by the dissipation of their extremely strong magnetic fields \citep{kfm+94}, rather than being rotationally-powered. Additionally, a unique trait of magnetars is their bursting properties, rarely shared with any other member of the neutron star population. These come in differing flavors. The short bursts are characterized by a $0.2$~s duration and hard X-ray energies reaching $10^{34}$~J \citep[e.g.,][]{collazzi}. On the other end, there are the most extreme events known as giant flares when up to $10^{40}$~J are released by a magnetar in about $0.5$~s in the hard X-ray, soft gamma-ray band \citep[e.g.,][]{palmer}. At times of bursting activity, magnetars may enter weeks to years-long episodes during which their persistent emission increases \citep{cotizelati}. These episodes are known as outbursts.

Currently, there are about 27 confirmed magnetars, with 3 potential candidates \citep{ok14}. These are overwhelmingly discovered during bursting periods (unlike canonical pulsars, which are mostly discovered via their persistent radio emission in large-scale surveys; e.g. \citet{mlc+01,slr+14,hww+21,mlk+24}, etc.). This is largely due to the lack of large field-of-view, sensitive X-ray surveys capable of identifying magnetar-like spin-periods without the need for deep follow-up observations.

Studying magnetars can also deepen our understanding of magnetic field evolution in neutron stars, providing crucial insights into the life cycles of these stars and the role of magnetic fields in their rotational evolution and outburst activities \citep{kb+17}. Additionally, magnetars are key to understanding the connection between neutron star populations, such as the possible evolutionary links between high-B pulsars, X-ray dim isolated neutron stars (XDINSs), and central compact objects (CCOs) which could help to solve the birth rate problem since the core-collapse supernova (CCSN) rate in the Galaxy measured today, $\beta_{\rm CCSN}$ is estimated to be $ 11.76 - 21.28$~kyr$^{-1}$ within a 68.3\% confidence interval \citep{rvc21}. This is smaller than the sum of all the estimates of birth rates of isolated neutron stars \citep{ge24}. Furthermore, recent detections of ultra long period magnetars (ULPMs) such as GLEAM-X118J162759.5-523504.3 might allow to increase the number of detected magnetars and even broaden the definition of what is a magnetar \citep{bwh+23}. Moreover, it is also plausible that a small fraction of milli-second magnetars produce gamma-ray bursts (GRBs) \citep{u92,dl98,kb10,2011MNRAS.413.2031M,mbg18,tbr+24} although their association as central engines of GRB jets is challenged on theoretical and observational grounds \citep{bl21,rtb+21,wdb+24}. They can also be responsible for fast radio bursts (FRB), the brightest events in the Galaxy, as evidenced by the association of FRB 20200428 with the known Galactic magnetar SGR 1935+2154  \citep{lkz20,brb+20,mbs+20}. While there is not yet a conclusive proof, various lines of evidence indicate that at least some, and perhaps all, extragalactic FRBs are associated with magnetars \citep{popov_2010,kumar+17,cc19,Gordon2023,Totani2023,BK2023,BK2024}. Their violent outbursts and persistent X-ray emission also provide constraints on magnetothermal evolution models, shedding light on the interplay between magnetic field decay, thermal conductivity, and neutron star cooling. In this sense, magnetars serve as natural laboratories for probing the complex interactions between magnetism, gravitation, and particle physics, offering valuable clues about the behavior of matter under extreme conditions. Also, understanding the magnetar population may lend clues on the links between them and events in the universe as well as newly discovered populations resembling magnetars.

Neutron star population synthesis is an efficient tool which allow to extract fundamental information about this population that would be unattainable otherwise, like their emission processes or the parameters characterizing the population (birth rate, magnetic decay timescale, mean magnetic field at birth, among others). In this type of study, neutron stars are generated from birth and evolved up to the present time. Once the sample is generated, detection criteria are applied to determine whether they can be observed with current and future instruments. Recently, \citet{grp+24,bec+24} applied simulation-based inference \citep[SBI, ][]{cbl+20} to pulsar population synthesis, demonstrating its effectiveness for such studies. By relying on simulations rather than analytical likelihoods (which may not be feasible or even possible to construct), SBI enables robust parameter estimation and model comparison.

Studies constraining the characteristic parameters of the population of magnetars have been done for instance, see \citet{gpm+15,bhv+19,jt+22}; where they constrained the mean magnetic field at birth, the birth rate, the magnetic decay timescale of magnetars. This paper combines magnetar evolution with the detection statistics while using simulation-based inference in a complete way that has not been included in previous studies. As such, finding these characteristic parameters will both improve models for magnetar evolution and provide insight for future searches. 

This paper is organized as follow: in Sec.~\ref{sec:popgen}, we describe how magnetar populations are generated, followed by an explanation of the magnetar evolution model in Sec.~\ref{sec:popevol}. Sec.~\ref{sec:obs} details the periodicity search conducted for this study, and Sec.~\ref{sec:detec} discusses the two types of detections considered in this work. Sec.~\ref{sec:sbi} provides an overview of the simulation-based inference method, while Sec.~\ref{sec:validation} focuses on validating the entire pipeline. The results are presented in Sec.~\ref{sec:results}, discussed in Sec.~\ref{sec:discussion}, and summarized in the conclusions in Sec.~\ref{sec:conclusions}.

\section{Generation of magnetars}
\label{sec:popgen}

We generate a population of magnetars whose ages are taken from a random uniform distribution between 0 and the maximum age of $5\times10^6$~yr (this is a conservative limit, as no observed magnetars have a characteristic age older than $3.6\times10^6$~yr and most are likely significantly younger). In each simulation, the number of sources is characterized by a birth spacing $X$, where $X$ is chosen as an integer number of years between the births of subsequent magnetars (i.e., $\frac{5\times10^6}{X}$ sources are drawn for each simulation). In this paper, we report this quantity as either the birth spacing $X$ in years or the birth rate $BR =  (1/X)\,{\rm yr}^{-1}$. This approach can be used to verify whether a constant birth rate is realistic, by comparing the results from the simulation to observation. Studies by \citet{bhv+19} and \citet{jt+22} found the birth rate of magnetars to be in the range $\rm 2.3-20\,kyr^{-1}$, corresponding to the birth spacing of $\rm 50 - 434\,yr$; the value of $X$ is drawn from this range. 

To describe the position of the magnetars in the Galaxy, we use the right-handed Galacto-centric coordinate system $(x,y,z)$ with the Galactic centre at its origin, $y$ increasing in the disc plane towards the location of the Sun, and $z$ increasing towards the direction of the north Galactic pole. The initial spatial distribution of the pulsars is given by the radial spread $\rho(R)$ from \citet{yk+04} and the spread in altitude $\rho_z(z)$ given by \citet{ok14}. 
\begin{subequations}
\begin{align}
\label{Yusu_eq}
\rho(R) &=A\left(\frac{R+R_1}{R_{\odot}+R_1}\right)^a \exp\left(-b\left(\frac{R-R_{\odot}}{R_{\odot}+R_1}\right)\right)  \\
\label{altitude_spread_eq}
\rho_z(z) & = \frac{e^{-|z+z_0|/h_e}}{h_e}       
\end{align}
\end{subequations}
where $R$ is the axial distance from the $z$-axis, and $z$ is the distance
from the Galactic disc. The numerical values for the constants are, $A = 37.6$~kpc$^{-2}$, $a = 1.64$, $b=4.0$, $R_1=0.55$~kpc, $h_e=30.7$~pc and $R_{\odot}=8.5$~kpc.
The purpose of choosing equation \eqref{Yusu_eq} to describe the radial spread is to put magnetars birth positions within the spiral arms of the Galaxy. More precisely, the Galactic spiral structure contains four arms with a logarithmic shape function, and the azimuthal coordinate $\phi$ is calculated as a function of the distance from the Galactic center,
\begin{equation}
\label{spiral_arms_eq}
\phi(R) = k \ln\left(\frac{R}{r_0}\right) +\phi_0 . 
\end{equation}
The values of the models describing each arm are given in Table~\ref{tabl_param_MW_spiral_arms}. These values were used by \citet{rgg+21} and taken from \citet{ymw+17} to match the shape of the arms of the Galaxy. The local arm is not modeled as it is far less dense than the four other arms. Each star has an equal probability to be in one of the four arms with angular coordinate $\phi$ calculated using \eqref{spiral_arms_eq} for a given $R$. Furthermore, the Galaxy is not static, and its arms are moving with an approximated period $T=250$~Myr \citep{ssm+19}. Because the Galaxy is rotating in the clockwise direction, the age of a magnetar will then dictate its angular position at birth. Following the same procedures described in \citet{rgg+21}, noise is added to both coordinates R and $\phi$ in order to avoid artificial features near the Galactic center. This correction factor is calculated via 

\begin{equation}
    \phi_{\rm corr}=\phi_{\rm rand} \, \exp(-0.35\,R)
\end{equation} 

where $\phi_{rand}$ is randomly drawn from a uniform distribution between $0$ and $2\,\pi$. Similarly, a correction factor for the radial distance $r_{corr}$ is drawn from a normal distribution with a mean of $1$ and a standard deviation $\sigma_{corr}=0.07$. These corrections are taken into account to compute the values $\phi$ and $R$ respectively:
\begin{equation}
    R_{\rm birth}=R \times r_{\rm corr}, 
\end{equation} 
and 
\begin{equation}
    \phi_{\rm birth}=\phi+\phi_{\rm corr} + \frac{2\,\pi\, t_{\rm age}}{T}
\end{equation} 
where $t_{\rm age}$ is the age of the magnetar. These coordinates are then converted to  Galacto-centric coordinates ($x,y$). \\
\begin{table}[h]
\caption{Parameters of the Milky Way Spiral Arm structure, adapted from \citet{ymw+17}.} 
\label{tabl_param_MW_spiral_arms} 
\centering 
\begin{tabular}{c c c c c} 
\hline\hline
Arm Number & Name & \(k \) & \( r_0 \) & \( \phi_0 \) \\
& & (rad) & (kpc) & (rad) \\
\hline 
1 & Norma & 4.95 & 3.35 & 0.77  \\ 
2 & Carina-Sagittarius & 5.46 & 3.56 & 3.82 \\ 
3 & Perseus & 5.77 & 3.71 & 2.09  \\ 
4 & Crux-Scutum & 5.37 & 3.67 & 5.76  \\ 
\hline 
\end{tabular}
\end{table}
Next, the rotational properties of each magnetar are produced, including the angle between rotation and magnetic axes (inclination angle) $\alpha_0$, the birth spin period $P_0$, and birth magnetic field $B_0$ (throughout this study, the term ``magnetic field" always denotes the surface magnetic field). The birth inclination angle $\alpha_0$ is assumed to follow an isotropic distribution generated from a uniform distribution $U\in[0,1]$ and given by $\alpha_0 = \arccos (2\,U-1)$.
In this work, $P_0$ and $B_0$ both follow log-normal distributions, as suggested from the results of a study of 56 young neutron stars by \citet{ifg+22}. These distributions are used in other magnetar population synthesis studies, including in \citet{jt+22} where a log-normal distribution is used for the spin period at birth and in \citet{rgp+15} where a log-normal distribution is used for the magnetic field at birth.

Explicitly, the probability distributions for $B_0$ and $P_0$ are given by
\begin{gather}
p(\log(B_0)) = \frac{1}{\sigma_b \sqrt{2\pi}} e^{-(\log B_0 - \log \mu_b)^2/(2\sigma_b^2)} \\
p(\log(P_0)) = \frac{1}{\sigma_p \sqrt{2\pi}} e^{-(\log P_0 - \log \mu_p)^2/(2\sigma_p^2)} \ .
\end{gather} 
where $\mu_{\rm b,p}$ and $\sigma_{\rm b,p}$ are the mean and the standard deviation of the log-normal distributions for magnetic field and period. 

Several mechanisms (asymmetry in the emission of neutrinos, hydrodynamic asymmetry during the collapse) have been studied to be at the origin of the asymmetry in the supernova explosion \citep{th75,c98,sp98,fkg+15}. These mechanisms seem to result in an alignment between the kick velocities at birth and the rotation axis of the pulsar, as suggested by \citet{r+07,nsk+13}. Therefore this alignment is taken into account in the simulation. Moreover, according to \citet{hll+05}, a Maxwellian distribution replicates well the observations of pulsar velocities (a description based on supernova (SN) simulations is provided in \citet{kmb+23}, but was not implemented here as it would have required progenitor parameters for the simulated neutron stars, which are not considered in this work). Additionally, here, we assume that magnetars originate from single-star progenitors, neglecting potential binary interactions that could affect their velocity distribution \citep{bp24}. Thus, in this work, pulsar velocities at birth are given by
\begin{equation} \label{eq:maxwellian_speed}
p(v) = \sqrt{\frac{2}{\pi}} \frac{v^2}{\sigma_v^3} \exp{\left(-\frac{v^2}{2\sigma_v^2}\right)}.
\end{equation}
The mean velocity of the distribution $\bar{v}$ is related to the standard deviation by $\bar{v} = \sigma_{v} \sqrt{{8}/{\pi}}$ with a standard deviation $\sigma_{v} = 265\,{\rm km/s}$ for pulsars. The velocity vector is then distributed along the unit vector $\vec{n}_{\Omega}$ of the rotation axis which is drawn from an isotropic distribution. The Cartesian coordinates of the unit rotation vector $\vec{n}_{\Omega}$ are $(\sin\theta_{n_{\Omega}} \cos\phi_{n_{\Omega}}, \sin\theta_{n_{\Omega}} \sin\phi_{n_{\Omega}}, \cos\theta_{n_{\Omega}})$.

\section{Evolution of magnetars}\label{sec:popevol}
\subsection{Evolution of the spin period, the magnetic field and the inclination angle}

The magnetar's initial period $P_0$, inclination angle $\alpha_0$, and magnetic field $B_0$ are evolved in time in a fully self-consistent way that takes into account spin-down losses and internal magnetic field dissipation within the neutron star. The neutron star period evolves according to the force-free magnetosphere model which takes into account the electric current and charge flowing inside the magnetosphere. The spin down luminosity~$\dot{E}$ is 
\begin{subequations}
\begin{align}    
\dot{E} & = \frac{dE_{\rm rot}}{dt} = - I \, \Omega \, \dot{\Omega} = L_{\rm ffe} \label{eq:spin_down_eq} \\ 
L_{\rm ffe} & = \frac{4\pi R^6 B^2 \Omega^4(1+\sin^2\alpha)}{\mu_0 c^3} \label{eq:value_of_spin_down_eq}
\end{align}
\end{subequations} 
where $I \approx 10^{38}\,{\rm kg\,m^2}$ is the neutron star moment of inertia, $\Omega = 2\pi / P$ the rotation frequency of the pulsar ($P$ is the spin period), $\dot{\Omega}$ is its time derivative, $R = 12\,{\rm km}$ is the typical radius of a neutron star as found by recent NICER observations \citep{rwb+19,blm+19}, $B$ is the magnetic field at the equator, $\alpha$ is the inclination angle, $\mu_0=4 \pi \times 10^{-7}\,{\rm H/m}$ is the vacuum permeability constant, and $c$ is the speed of light. The term $L_{\rm ffe}$ for the spin down luminosity was given by \citet{s+06} and \citet{p+12}. Equation \eqref{eq:spin_down_eq} clearly shows the correlation between the obliquity, the magnetic field, and the rotation frequency. Combining equation \eqref{eq:spin_down_eq} and \eqref{eq:value_of_spin_down_eq} leads to 
\begin{subequations}
\begin{align}    
\dot{\Omega} & = -K_{\rm ffe} \ \Omega^n \label{eq:general_omegadot} \\ 
K_{\rm ffe} & = \frac{4\pi R^6 B^2(1+\sin^2\alpha)}{\mu_0 c^3 I} \label{eq:kffe},
\end{align}
\end{subequations} 
where $n$ is the braking index ($n=3$ for magnetic dipole radiation).

The most speculative part of our work regards the evolution of the inclination angle, as it is still a hotly-debated issue. From a theoretical point of view, the study of \citet{ptl+14} suggests that the force free model allows for an evolution of $\alpha$ different from the exponential decay, that applies when considering a neutron star in vacuum (as applied to observed data by e.g. \citealt{jt+22}). Another theory points out that the inclination angle should tend to $\pi/2$ to minimize energy loss \citep{bgi93,bn07,bip13}. However, from an observational point of view, there is no clear evidence in favor of one of these approaches; even when there is a tendency toward alignment with increasing age, there is a large spread \citep{tm98,wj08}. Finally, the free precession of neutron stars studied in \citet{m00,apt15,ip20} shows that there should be oscillations but long-term decay in the evolution of $\alpha$, which could explain why we observe neutron stars that seem to have an increasing inclination angle \citep{ljg+15}. Since we are considering the force-free model in this work, we will use the integral of motion between $\Omega$ and $\alpha$ found in \citet{ptl+14} which reads
\begin{equation} \label{eq:integralofmotion}
\Omega \frac{\cos^2\alpha}{\sin\alpha} = \Omega_0 \frac{\cos^2\alpha_0}{\sin\alpha_0}
\end{equation}
where quantities with subscript 0 indicate their initial value and those without a subscript are the value at present time. 

The evolution of neutron stars' magnetic fields is also debated. Few pulsar population syntheses claim that the magnetic field of neutron stars was not evolving \citep{bwh+92,fgk06}. However, as discussed earlier, magnetars cannot be powered by spin-down. They are therefore highly likely to be powered by magnetic field decay. Several phenomena strengthen the idea that the magnetic field should decay: ohmic dissipation in the crust, ambipolar diffusion in the core, vortex movements in the crust, internal turbulence and the loss of energy via radiation emission \citep{caa+04,dgp12}. In the specific case of magnetars, the X-rays could not be powered by spin-down and therefore be powered magnetically, implying the decay of the magnetic energy reservoir \citep{kb+17}. In addition, a recent study which takes into account ohmic heating \citep{ip24}, period derivative measurements and spectral information about the neutron star RX J0720.4-3125, suggested that its magnetic field was decaying rapidly. Furthermore, it has been shown in other and more recent pulsar population studies that a decaying magnetic field better reproduces the observed pulsar/magnetar population \citep{bhv+19,jt+22,spm+24}.
Therefore, we prescribe a magnetic field decay according to a power law , i.e. 
\begin{equation} \label{eq:Bfield}
B(t) = B_0\left(1+\frac{\alpha_d t}{\tau_d}\right)^{-1/\alpha_d}
\end{equation}
where $\alpha_d$ is a constant parameter controlling the speed of the magnetic field decay and $\tau_d$ the typical decay timescale.  

In line with the spin down luminosity, the inclination angle satisfies another evolution equation that after integration was found by \citet{ptl+14} for a spherically symmetric neutron star with a constant magnetic field.
With our decaying prescription the inclination angle $\alpha$ is found by solving for the root of
\begin{multline}
\scriptsize
\label{eq:incl_angle}
\ln(\sin \alpha_0) + \frac{1}{2\sin^2\alpha_0} + \frac{K \Omega_0^2 \tau_d B_0^2}{\alpha_d - 2} \frac{\cos^4\alpha_0}{\sin^2\alpha_0} \\
\times \left[\left(1+\frac{\alpha_d t}{\tau_d}\right)^{1-2/\alpha_d}-1\right] = \ln(\sin \alpha) + \frac{1}{2\sin^2\alpha}
\end{multline}
where $t$ is the age of the pulsar and $K={4 \pi R^6}/{I \mu_0\,c^3}$. The typical decay timescale for a mean magnetic field of $10^{10}$~T is $3 \times 10^4\,{\rm yr}$ in \citet{v+13} for instance. However as we will show in Section \S\ref{sec:sbi}, we do not assume that $\tau_d$ depends on magnetic field, instead we use statistical inference to determine $\tau_d$. 

\subsection{Description of the Galactic potential}

Finally, the motion of the magnetar within the Galactic potential is calculated. Each magnetar evolves in the gravitational potential $\Phi$ subject to an acceleration $\boldsymbol{\ddot{x}}$ according to 
\begin{equation} \label{move_equation}
\boldsymbol{\ddot{x}} = -\boldsymbol{\nabla} \Phi. \
\end{equation}
Equation~\eqref{move_equation} is integrated numerically thanks to a Position Extended Forest Ruth-Like (PEFRL) algorithm \citep{omf+02}, a fourth order integration scheme (see Appendix A of \citealt{spm+24} for a detailed analysis of the PEFRL scheme and its accuracy). We now discuss the galactic potential model used.

The Galaxy is divided in four distinct regions with different mass distributions and associated gravitational potentials. The four potentials are: the bulge $\Phi_{b}$, the disk $\Phi_{d}$, the dark matter halo $\Phi_{h}$ and the nucleus $\Phi_{n}$. The total potential of the Milky Way $\Phi_{tot}$ is the sum of these potentials:
\begin{equation} \label{tot_pot}
\Phi_{tot} = \Phi_{b} + \Phi_{d} + \Phi_{h} + \Phi_{n} .
\end{equation} 
The expressions for these potentials are taken from \citet{bb+21} with parameters given in Table \ref{tabl_const_pot}. The nucleus mass was found in \citet{b+15}. 

\input{tables/constants}

The potential for the bulge $\Phi_b$ and for the disk $\Phi_d$ both have the forms proposed by \citet{mn+75} which are typically used for models of the gravitational potential of the Milky Way. They read
\begin{equation} \label{potMN}
\Phi_i(R,z) = -\frac{GM_i}{\left[R^2+\left(a_i+\sqrt{z^2+b_i^2}\right)^2\right]^{1/2}}
\end{equation}
where $R^2=x^2+y^2$, and $i=b$ is for the bulge and $i=d$ is for the disk. In those formula $R$ depends on the coordinates $x$ and $y$, $a_i$ and $b_i$ are the scale parameters of the components in kpc, $M_i$ is the mass (for the disk or the bulge) and $G$ is the gravitational constant. The values of these constants can be found in Table \ref{tabl_const_pot}. 

The potential for the dark matter halo $\Phi_h$, according to the frequently used potential of \citet{nfw+97}, is expressed as
\begin{equation} \label{NFW_eq}
\Phi_h(r)=-\frac{GM_h}{r} \ \ln\left(1+\frac{r}{a_h}\right)
\end{equation}
where $M_h$ is the mass of the halo, $r=x^2+y^2+z^2$, and $a_h$ is the length scale whose values are found in Table~\ref{tabl_const_pot}. 

Finally for the last part of the Galaxy, the nucleus of the Milky Way is simply represented by a Keplerian potential $\Phi_{n}$ such as
\begin{equation} \label{kep_pot}
\Phi_{n}(r) = -\frac{GM_{n}}{r}
\end{equation} 
where $M_{n}$ is the mass of the nucleus. 

\subsection{X-ray emission mechanism} \label{subsec:X_ray_subsection}

X-ray emission from magnetars is usually associated with resonant Compton scattering (RCS) or thermal emission \citep{rzt+08,zrt+09,gpm+15,rgp+15}. For the former, electrons in the magnetosphere absorb nearby photons and re-emit them as X-rays afterwards. In this work, we focus exclusively on the dipolar magnetic energy losses ($\dot{E}_B$), which account for a fraction of the total magnetic energy losses ($0.1 \leq f_{dip} \leq 0.3$). Most of the magnetic energy is stored in the toroidal component of the magnetic field, which is not considered here. The persistent X-ray luminosity of the star appears to represent a fraction $f_{per} < 1$ of the total magnetic energy losses \citep{bwt+24}. Based on current understanding, it is reasonable to assume that $f_{dip} = f_{per} = 1/3$, leading to the following estimation for the persistent X-ray luminosity of a simulated magnetar: $L_X \approx \dot{E}_B$. Furthermore, the dipole magnetic energy ($E_B$) and the dipole magnetic energy losses can be written as 
\begin{subequations}
\begin{align}
E_B = \frac{2 \pi R^3 B^2}{3 \mu_0}, \label{eq:E_dip} \\
\dot{E}_B = \frac{4 \pi R^3 B \dot{B}}{3 \mu_0} \label{eq:dot_B} 
\end{align}
\end{subequations}
where $R$ is the radius of the neutron star (12 km in our simulations), $\mu_0$ is the permeability constant and $B$ is the magnetic field. This luminosity is then converted into an unabsorbed flux 
\begin{equation}
S = \frac{L_X}{4 \pi d^2}
\end{equation}
where $d$ is the distance of the magnetar to us. By assuming blackbody emission and under an assumption of the magnetar's emitting surface fraction in X-rays, the surface temperature can be calculated as such in the simulation:
\begin{equation} \label{eq:bbem}
T = \left(\frac{L_X}{\xi\times4 \pi R^2 \sigma}\right)^{1/4}
\end{equation}
where $\xi$ is the emitting surface fraction in X-rays, $\sigma \approx 5.67\times10^{-8} $\,W\,m$^{-2}$\,K$^{-4}$ is the Stefan-Boltzmann constant and $T$ is the temperature. This computation is very complicated observationally because of the deviation of the emission from a pure blackbody spectrum, the complex emission geometry on the surface, and the poorly known distances to the sources. Nonetheless, we explored varying estimates of the emitting surface fraction for our simulated magnetars which were considered detected in the process. This is discussed further in \S\ref{sec:results}.

\section{Observations and Data Reduction}
\label{sec:obs}

A crucial component of the simulation process is the comparison of the summary statistics measured from each simulated population to equivalent measurements from the ``true" population of magnetars. One of these statistics comes from a search for new magnetars within the observed XMM-Newton sky; in particular, the number of magnetars discovered by an untargeted search. As there is no complete survey of the Galactic plane in X-rays, archival XMM data were utilized as an ad hoc survey; a periodicity search was conducted on point sources identified in the 4XMM-DR13 catalog \citep{wct+20}. Unlike modern surveys for pulsars, where large regions of the Galaxy are uniformly searched to identify new sources and place constraints on the entire population, the XMM observations cover a far smaller region of the sky. Additionally, these observations are primarily targeted observations of previously-identified sources as opposed to ``blind" searches. For these reasons, the effective ``survey" used is very uneven and biased. Nevertheless, searching for sources with well-understood sky coverage can begin to inform population and evolution models for the Galactic magnetar population. 

Our search is based on 4XMM-DR13 catalog. We exclude all known bright sources (e.g., the Crab pulsar) and only consider sources that are within $\pm5^\circ$ of the Galactic plane; magnetars age rapidly and are not expected to be detected beyond such a high Galactic latitude. Also, we only consider sources with total number of counts $\geq 200$. Finally, we exclude observations in small window and timing mode were omitted.

These restrictions reduced the number of sources to search to 26073 sources contained in 1584 observations. For each candidate source, we performed data reduction using tools in the \texttt{XSPEC} \citep{a96} and XMM-Newton Science
Analysis System (XMM-SAS) version 21.0.0 tools. First, the raw observation data in which the candidate was observed is downloaded. Using the full 4XMM catalog, all known sources (aside from the candidate of interest) are extracted from the dataset using the ``EXTENT" parameter as the extraction radius to create a ``Swiss cheese" image; this is the background for each observation. High energy ($\geq10\,{\rm keV}$) background sources associated with proton flaring are also removed. The 4XMM catalog provides observed count rates in several energy bands: $0.2-0.5\,{\rm keV}$ (band 1), $0.5-1\,{\rm keV}$ (band 2), $1-2\,{\rm keV}$ (band 3), $2-4.5\,{\rm keV}$ (band 4), $4.5-12\,{\rm keV}$ (band 5), and the full band (band 8). To determine the best band for the search, the signal-to-noise (S/N) for each band and all contiguous combinations were calculated using the following expression:

\begin{equation} \label{eq:sn}
    \rm S/N = \frac{\sum_{i} r_i }{\sum_{i} \sigma_i} \sqrt{\tau_{\rm exp}}
\end{equation}
where ${\rm r_i}$ is the rate in band $\rm i$, ${\rm \sigma_i}$ is the uncertainty on the rate in band $\rm i$, and $\tau_{\rm exp}$ is the duration of the exposure. The band with the highest S/N was used for the search, and an image was created using this energy band. Next, a King function was fit to the source to find an optimal radius of extraction:

\begin{equation}
       \rm K(r) = A\,\left(1+\left(\frac{r}{r_0}\right)^2\right)^{-\beta} + DC
\end{equation}
where A is a scaling factor, $\rm r_0$ is a characteristic angular radius, $\beta$ is the index of radial decay, and $\rm DC$ is a constant offset from zero. These four parameters were fit to the radial intensity curve produced by the \texttt{eradial} routine in XMM-SAS. Using this fit, the optimal extraction radius was chosen to include 99\% of the area under the curve, and a time series dataset was created from the background-subtracted source region. This created our final "cleaned event list" for each source.

With the time series in hand, a search for periodicity was conducted using an H-test \citep{db10}. The search spans frequencies between $\rm50\,mHz$ and $\rm2\,Hz$; step sizes are chosen to equal the inverse of the observation duration or 5e-5, whichever is larger. This limit is chosen to prevent excessive computation time for observations lasting over $\rm20\,ks$. All calculated H-statistics are saved, and significant detections are highlighted. For these candidates, spectral files are also created for additional confirmation. To choose a significance threshold, the same H-test was conducted on simulated time series data containing only noise, and all resulting H-statistic measurements were binned in a histogram. From this, a CDF was measured, and the value of the H-statistic below which 99.999\% of samples were included was determined to be 47. Using the PDF as measured by \cite{db10}, 47 sits even closer to 100\%, and so it was accepted as a very conservative limit. The search turned up no new sources, though 14 previously-known magnetars were detected.

\section{Simulated detection} \label{sec:detec}
\subsection{Magnetar outbursts}

Given that all the known magnetars were discovered through outbursts, modeling this phenomenon in our simulation was very important to match the results to the observed population. The process used here is very similar to the one in \citet{bhv+19}, where it is assumed that the burst energy distribution per magnetar is dominated by the most energetic bursts ($dN/dE \propto E^{-\gamma} $ with $1.4 \leq \gamma \leq 1.8$ \citep{gwk+99,gwk+00}, and the total energy released is $\propto E^2 dN/dE$). In addition, the energy released during a GF is on the order of $10^{37}\,{\rm J}$, and the burst rate scales with the normalized magnetic energy loss $\dot{E}_B/E_{B,0}$. We assume 
\begin{equation} \label{eq:ratio_Bdot_B}
\frac{\dot{E}_B(t)}{E_{B,0}} = \frac{2 B \dot{B}}{B_0^2}.
\end{equation}
For a magnetar with age $5\times10^3\,{\rm yr}$ and initial magnetic field of $3\times10^{10}\,{\rm T}$, giant flares are expected to occur at a rate of approximately $\rm 10\,kyr^{-1}$ for a typical energy of $10^{37}$~J. 
Given that hard X-ray burst-monitors have been active for approximately 50 years, for each magnetar in the simulation, we determine whether it has experienced an outburst within the last 50 years using its burst rate. For these simulations, we assume that all bursts in this timeframe are detected, although it might slightly overestimate the number of detections; no burst detection threshold is applied.

\subsection{Convolution with XMM-Newton survey and periodicity search}
After producing a simulated population of Galactic magnetars, we used the XMM coverage of the Galactic plane to determine which (if any) of the sources would be detected by periodicity search. The source positions were matched to the catalog of XMM observations, and sources that fall within the observation footprint were retained. In cases where multiple observations overlap the same position, we used the longest available. Vignetting in XMM observations reduces the effective exposure away from the boresight, and some pixels are corrupted. This information is encoded in the exposure maps, which we produced for each XMM observation using the \texttt{expmap} routine in XMM-SAS. From these, a precise exposure was found for each source. Using the distances to each simulated source and its sky position, we estimated the expected column density of hydrogen ($\rm N_H$) using the 3D-NH tool\footnote{\url{http://astro.uni-tuebingen.de/nh3d/nhtool}} \citep{d24}. This tool utilizes X-ray, visible, and radio data to estimate absorption along a given line of sight for a provided distance. At this point, we determined the optimal temperature band for a search by simulating source and background count rates in each 4XMM-DR13 band by modeling an absorbed blackbody in \texttt{XSPEC} and choosing the band with the highest S/N, per equation \ref{eq:sn}. 

With the exposure, optimal band, and count rates in hand, we simulated a time series for each candidate source. We drew pulsed fractions from an empirical PDF derived from a double-Gaussian fit to known magnetar pulse fractions \citep{hnh19}. A sinusoidal pulse profile was produced with individual phase bins that are required to be greater or equal to the XMM sampling time. Source counts were distributed across each rotation of the magnetar according to its pulse fraction, and background counts are distributed according to Poisson noise. As was done in the ``real" search, we conducted an H-test on the lightcurve with bin sizes corresponding to the inverse of the exposure or 20\,$\mu$Hz, whichever is larger. Because the source spin frequency is known and the simulation is intended to determine simply whether or not the source is detected, only the frequency bins within 5\% of the source frequency are used in the search to save processing time. Any sources detected with H-statistics above 47 (the search threshold) at their spin frequency are considered detections via periodicity search.

\section{Simulation-based inference} \label{sec:sbi}
\subsection{Concept}
To constrain the parameters which characterize the Galactic magnetar population, simulation-based inference (SBI; \citealt{cbl+20}) was utilized. Consider a model $M$ with a vector of parameters $\theta$ which produces data $D$. In traditional statistical inference, we assume that the data generating process $p(D | \theta)$, also called the likelihood, can be evaluated analytically. We can then find the probability of the model parameters, also called the posterior probability, via Bayes theorem:

\begin{equation}
p(\vec{\theta} | \vec{D}) = \frac{p(\vec{D} | \vec{\theta})p(\vec{\theta})}{p(\vec{D})} \; ,
\end{equation}

where $p(\vec{\theta})$ denotes the prior probability distribution of the parameters, and $p(\vec{D})$ is called the marginal likelihood. In parameter estimation problems, where we are interested primarily in the \textit{shape} of the distribution, we generally only need to consider $p(\vec{\theta} | \vec{D}) \propto p(\vec{D} | \vec{\theta})p(\vec{\theta})$, since $p(\vec{D})$ is a normalization constant and does not depend on the parameters. 

Evaluating the posterior requires an analytical form for the likelihood. This may not be available in practice, for example, when the measurement process introduces complex effects that are not easily parametrized, or when the output of the data generating process is inherently stochastic. In these cases, it may be possible to \textit{simulate} data sets that are somewhat faithful representations of the real data, but we cannot easily relate the model to the observed data via the likelihood.

In this context, SBI provides principled methods for relating simulated data to the underlying model parameters such that (unbiased) inference becomes possible. In SBI, we use a large number of pairs of parameters $\theta$ and simulated data $D$ to learn a mapping between them to approximate the likelihood (Neural Likelihood Estimation, \citealt{psm+18,acf+19}), the posterior (Neural Posterior Estimation, \citealt{pm+16,msc+22,vra+23,grp+24}) or the ratio of likelihood to marginal likelihood (Neural Ratio Estimation, \citealt{hbl+19,bam+23}). Flexible neural network-based density estimators called \textit{Normalizing Flows}, designed to approximate complex probability distributions, are often used to learn the mapping between parameters and data. 

Because the data is often high-dimensional and noisy, it can be much easier to first compute \textit{summary statistics} $\vec{x}$ of the raw (simulated) data: these summaries can be engineered to explicitly encode properties of the data that are informative with respect to the parameters, or they can be learned automatically, often again using neural networks \citep{grp+24}. 

Here, we use Neural Posterior Estimation (NPE), since we are particularly interested in the posterior distributions from our observed data. We generate training data from flat priors for all parameters (see Section \ref{sec:sim_setup} and Table \ref{tab:inparam} for details), compute summary statistics (explained in more detail in Section \ref{sec:sim_setup}), and  train a Normalizing flow to approximate the posterior probability of the population model parameters given the observed sample of magnetars. We use the python package sbi\footnote{\url{https://github.com/sbi-dev/sbi}} \citep{sbi-soft} for neural network setup and training.

\subsection{Simulation setup}\label{sec:sim_setup}

The characterizing model parameters of the Galactic magnetars population are : $\vec{\theta} = \{\sigma_b,\mu_b,\sigma_p,\mu_p,\alpha_d,\tau_d,BR\}$ (see \S\ref{sec:popgen} and \S\ref{sec:popevol} for descriptions). The ranges for the flat priors on the model parameters are given in Table \ref{tab:inparam}. Note that $\alpha_d$ can be negative, and by the definition of the magnetic field decay in equation \eqref{eq:Bfield}, this allows for a simulated magnetar to reach a negative magnetic field. Because this is non-physical, a magnetar which reaches a negative magnetic field is ruled out for detection. These ranges were chosen to explore a similar, and an even bigger parameter space than what was done in \citet{bhv+19,jt+22}.

\input{tables/input_parameter_ranges}

Since the parameter space is relatively large, the Latin hypercube sampling (LHS) method \citep{LHS_paper} is used to explore it most efficiently. This method divides the parameter space into a specified number of intervals (depending on how many simulations are made) and samples the model parameters such that only one value is drawn per interval. Therefore, this method ensures the values chosen for the model parameters in different simulations are representative of the whole space. 

The simulations each produce 8 summary statistics (5 vectors and 3 scalars) to characterize the population. In order to compare the simulated population with the observed population (which has been discovered almost entirely through bursts), the first six statistics are obtained from the magnetars which experienced at least one outburst episode in the last 50 years when burst monitoring has been active. Five of these statistics are feature vectors from the following density maps (which are 2D histograms of size 32 $\times$ 32):
\begin{itemize}
    \item $P-\dot{P}$, 
    \item $P-L_X$, 
    \item $P-g_b$, 
    \item $\dot{P}-L_X$, and 
    \item $\dot{P}-g_b$. 
\end{itemize}
The feature vectors from the density maps of the simulated magnetars detected through outburst, are obtained after the density maps were passed to a CNN which is detailed further below in this subsection. The sixth population statistic is the slope of the log N - log S curve, dubbed $K$. 

The final two summary statistics are related to detection, and are therefore compared with the results of the search conducted in this project (as well as the observational history of magnetars): 
\begin{itemize}
    \item $n_{detect}$ : the number of magnetars detected by periodicity search in XMM-Newton data, and
    \item $n_{outburst}$ : the total number of outburst episodes of magnetars in the last 50 years. 
\end{itemize}

These 8 outputs of the simulations are then compared to the same characteristics of the observed data in order to make the best possible match. 

The physical properties of magnetars (e.g. $P$ and $\dot{P}$ are often found to be correlated, mirroring the underlying physical relationship (e.g.~of the aging process of magnetars). Consequently, simple summaries of the data such as the mean and the standard deviation of the simulated population are insufficient to capture important relationships with respect to the input parameters.
In order to characterize the quantities and their correlation, that is why we generated density maps of $P-\dot{P}$, $P-L_X$, $P-g_b$, $\dot{P}-L_X$ and $\dot{P} - g_b$ and then a convolutional neural network (CNN) is used to generate informative features from these maps (each type of map will have its own dedicated CNN for training). These features are 1-D vectors, that represent the latent features extracted by the CNNs which can be parsed by the density estimator easily. 

\begin{figure*}
\centering
\resizebox{\hsize}{!}{\includegraphics{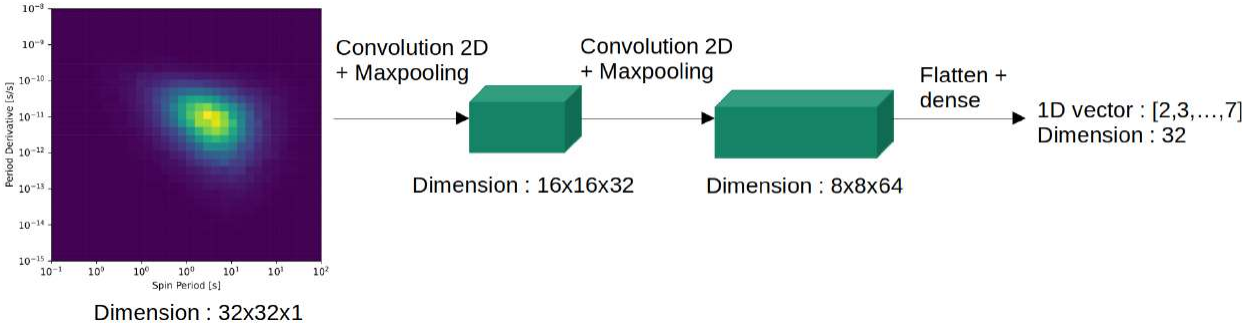}}
\caption{Convolutional neural network architecture used for this work.}
\label{CNN_arch}
\end{figure*}

The CNNs used in this work have the following architecture (see Fig.~\ref{CNN_arch}; \citealt{grp+24} proposed a similar CNN architecture): 
\begin{itemize}
    \item Two-dimensional convolution layer with kernel size 3 × 3, 3 input channels, 32 output channels, stride 1, padding "same".
    \item Two-dimensional Max pooling layer with size 2×2, stride 2, no padding.
    \item Two-dimensional convolution layer with kernel size 3 × 3, 32 input channels, 64 output channels, stride 1, padding "same".
    \item Two-dimensional Max pooling layer with size 2×2, stride 2, no padding. 
    \item Fully connected linear layer with the flattened output from the second pooling layer as input and 32 output neurons encoding the latent representation. 

\end{itemize}

After each convolution and the fully connected layer, a rectified linear unit (ReLU) activation function is applied. 

The summary statistics (the 5 vectors from the density maps and the 3 scalars) of the simulation were then passed to the default density estimator of SBI, a Masked Autoregressive Flow (MAF; \citealt{ppm17}), a type of normalizing flow that models complex distributions through a sequence of autoregressive transformations. The model was trained by minimizing the Kullback-Leibler (KL) divergence \citep{kl+51}, ensuring that the learned distribution closely matches the true posterior. This approach is particularly well-suited for Neural Posterior Estimation (NPE), as it enables flexible and efficient density estimation. 

In this work, 10490 simulations were employed to train the density estimator of SBI for parameter inference. This number was chosen based on a specific criteria: following the recommendation from other SBI users, who suggest using at least $1000\times n$ simulations, where $n$ is the number of parameters. Additionally, due to the complexity of simulating magnetars, the number of simulations for inference was extended to 10000 to ensure confidence in the results. 

\section{Model validation} \label{sec:validation}

\begin{figure*}
\centering
\resizebox{\hsize}{!}{\includegraphics{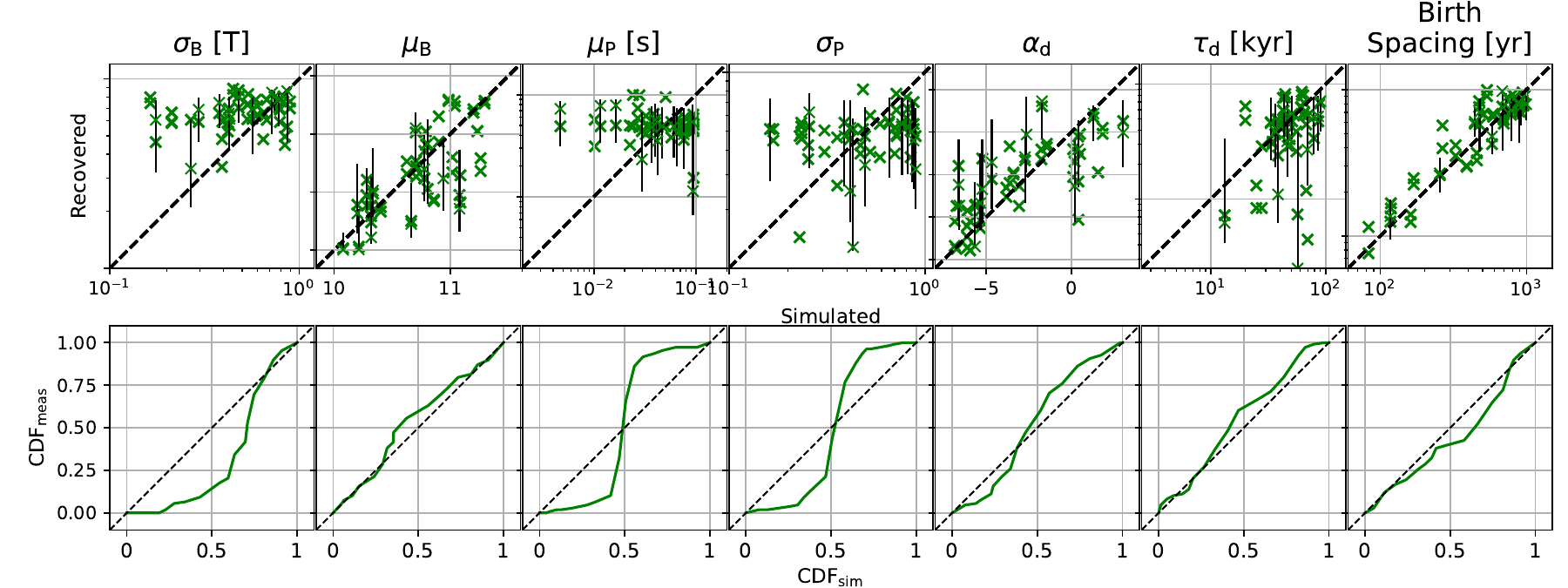}}
\caption{Validation plot for the SBI pipeline. In the upper panels, input values are plotted against their recovered counterparts directly with 68\% credible interval error bars. The lower panels compare the CDFs of both the input and output parameters directly. If the pipeline has successfully recovered parameters, these lower plots should track the dotted line (unity); deviations indicate regions where the parameters are not recovered accurately.}
\label{fig:validation}
\end{figure*}

\begin{figure}
    \centering
    \resizebox{\hsize}{!}{\includegraphics{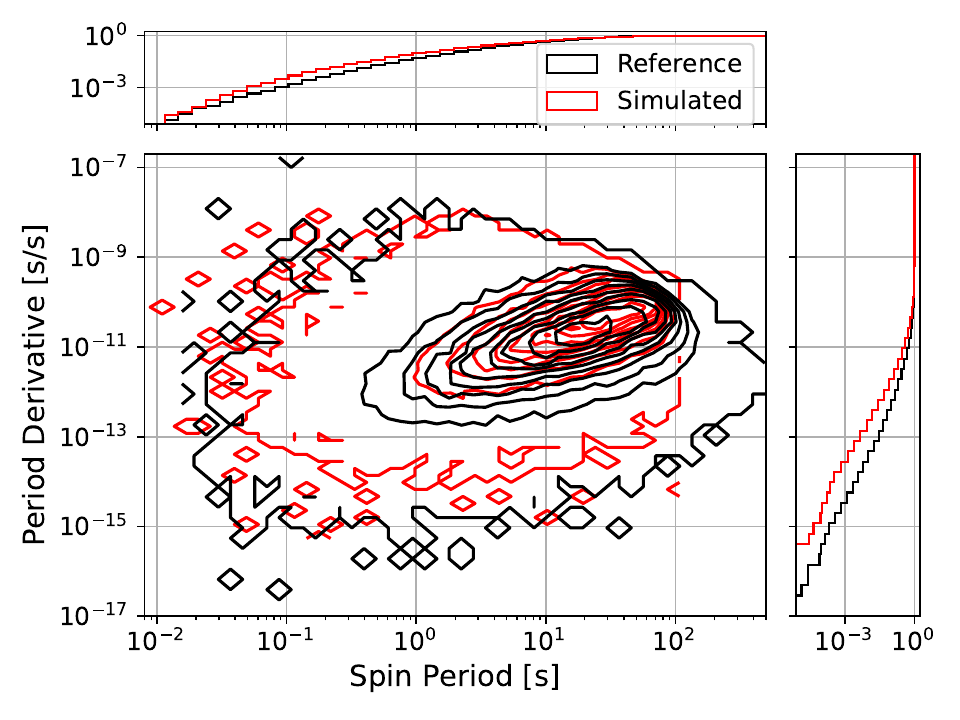}}
    \caption{$P-\dot{P}$ diagrams following pipeline validation checks. The main panel includes contour maps for two collections of 5000 populations: the black contours correspond to the simulation constructed using a known set of parameters drawn from the prior ($\vec{\theta_{\rm in}}$), and the red are the populations generated from parameters drawn from the posteriors following a full run of the SBI pipeline ($\vec{\theta_{\rm out}}$). Along the top and right side are the CDFs for $P$ and $\dot{P}$ for both populations.}
    \label{fig:validation_ppdot}
\end{figure}

We performed validation checks to test the  robustness of our SBI pipeline. First, a population of magnetars was simulated following the technique described in \S\ref{sec:popgen} and \S\ref{sec:popevol} with a known set of model parameters $\vec{\theta_{\rm in}}$. This dataset replaced the ``true" magnetar population, and the SBI pipeline was run under this assumption to check whether the model parameters obtained with SBI $\vec{\theta}_{\rm out}$ match the known set of model parameters $\vec{\theta_{\rm in}}$ and we repeated this process many times. The results from these tests are given in Figure~\ref{fig:validation} which shows that the ground truth values were largely recovered by the SBI process except for the birth period parameters ($\mu_{\rm p}$ and $\sigma_{\rm p}$). Birth rate (1/$X$) and the field decay parameters ($\alpha_{\rm d}$ and $\tau_{\rm d}$) are the parameters that are better recovered, with the majority of results matching parameters within the 95\% credible interval. To further validate the results, a second comparison is performed using 5000 $P-\dot{P}$ diagrams from populations where contours were created for bursting sources. These populations were generated using both the known set of model parameters $\vec{\theta_{\rm in}}$ and the model parameters inferred from the posteriors obtained with SBI. These are overlaid in Figure~\ref{fig:validation_ppdot}, and the CDFs of both distributions are given as well. It can be seen here that the contour of the reference population (generated with $\vec{\theta_{\rm in}}$) is quite well reconstructed from the distributions of parameters found by SBI, validating the pipeline's ability to find the parameters. It shows the pipeline's insensitivity to birth period, suggesting that the current magnetar population is insensitive to the parameters $\mu_p$ and $\sigma_p$ (as reported in other works, e.g. \citealt{gpm+15}, \citealt{bhv+19}, and \citealt{jt+22}). This is most likely due to the fast decay of the magnetar spins and magnetic fields, as a magnetar will be detected with ages comparable to $\tau_d$ but the timescale of modification of $P$ is much shorter than $\tau_d$. Because of this, the birth spins are very hard to determine with a small sample size; with only 30 known magnetars, we are unable to sufficiently probe the initial spin period. We further detail our validation steps in \hyperref[sec:AppA]{Appendix}.

\section{Results} \label{sec:results}

\begin{figure*}[t]
\centering
\includegraphics[width=\textwidth,height=0.5\textheight,keepaspectratio]{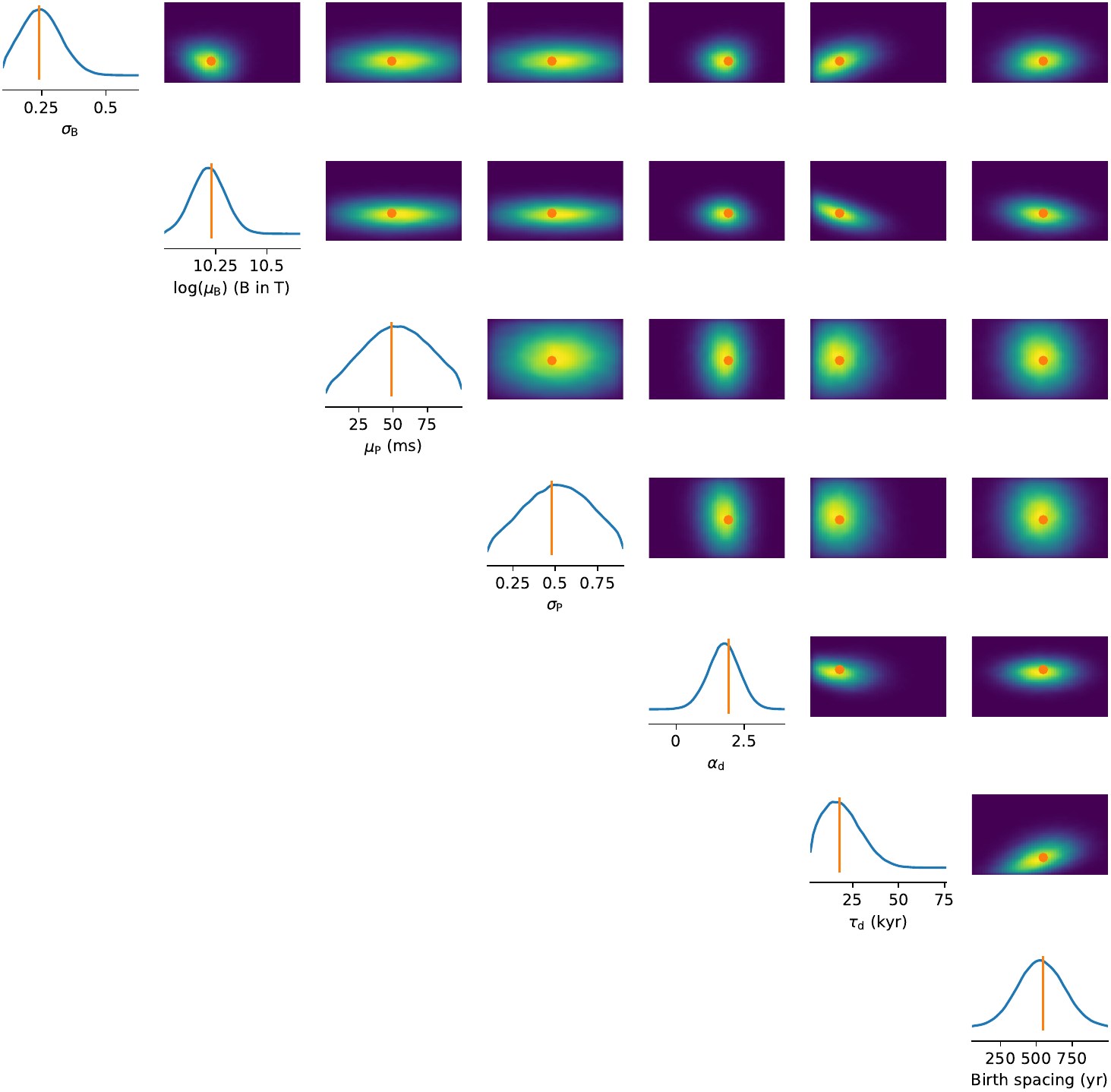}
\caption{Corner plot of results obtained with SBI. Vertical yellow lines indicate the best estimate.}
\label{cornerplot_obs}
\end{figure*}
\begin{figure}[h]
\centering
\resizebox{\hsize}{!}{\includegraphics{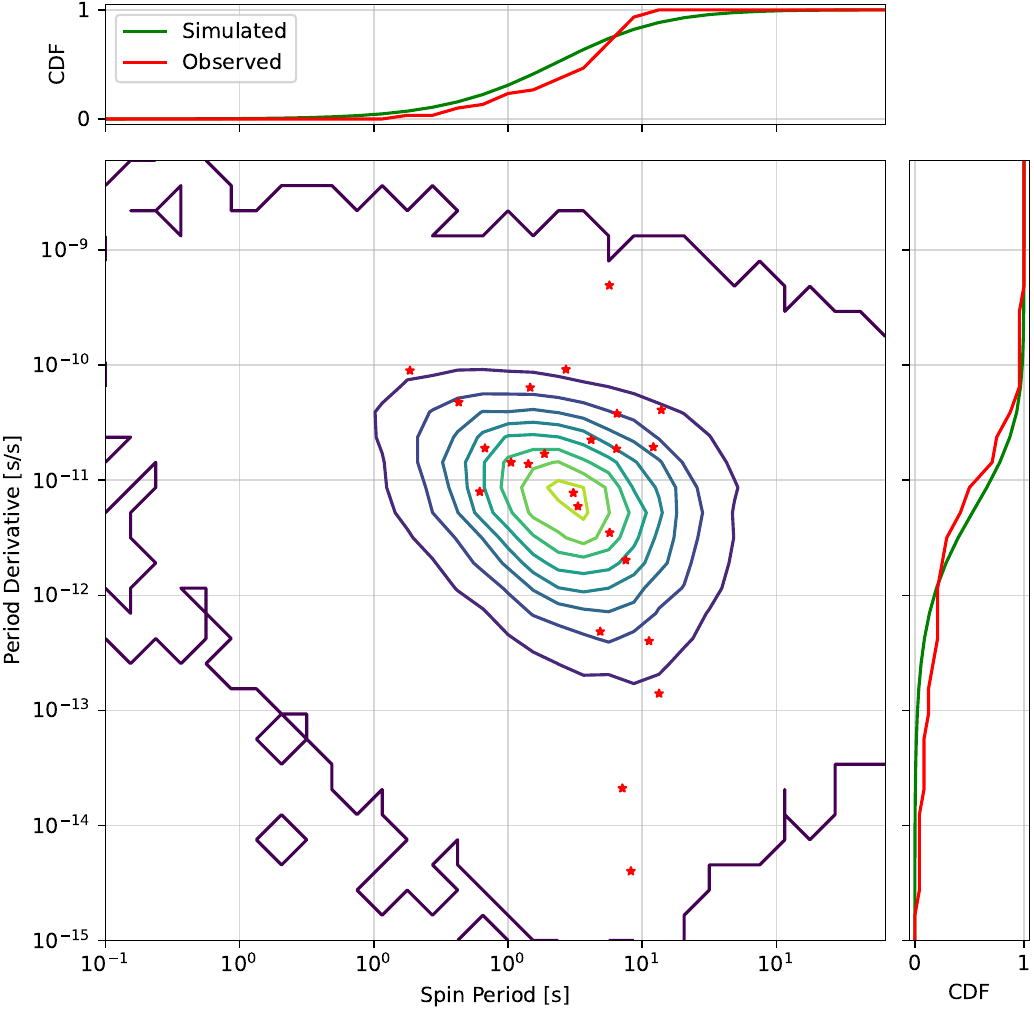}}
\caption{Contour $P-\dot{P}$ diagram. To produce the contours, 5000 simulations were obtained using the posterior distributions of the parameters obtained at the end of the SBI. Only sources which have undergone an outburst in the past 50 years are included, as this is the subpopulation we assume to have detected. Plotted in red are the observed Galactic magnetars. Along the side and top of the plot are the CDFs of each distribution for comparison.}
\label{contour_map_ppdot_result}
\end{figure}

We applied the SBI pipeline to the observed magnetar population. The posterior distributions for the model parameters are shown in Fig.\ref{cornerplot_obs}. The best estimates are presented in Table~\ref{tab:results} with their 95\% credible intervals. We do not report the values found for $\mu_p$ as the validation checks suggest that this parameter is not well recovered with the available data. However, while we do not show the tests here, using the best estimates from Table~\ref{tab:results} with different $\mu_p$ and $\sigma_p$ made clear that 1~s for $\mu_p$ and 0.9 for $\sigma_p$ are appropriate upper limits, as models far from these values lead to $P-\dot{P}$ diagrams where simulations and observations no longer align. Note that the measured birth spacing reported in Table~\ref{tab:results} corresponds to birth rates ${\rm BR}=1.8^{+2.6}_{-0.6}$~kyr$^{-1}$; our result provides a significantly tighter constraint compared to the bounds of $2.3$--$20\,{\rm kyr}^{-1}$ reported in the previous study of \citet{bhv+19}.
\begin{table}[h]
\caption{Results of the SBI pipeline. Uncertainties indicate the 95\% credible interval.}
\label{tab:results}
\centering
\begin{tabular}{c c}
\hline\hline 
Parameter [unit] & Value \\
\hline 
$\sigma_b$     & $0.2^{+0.2}_{-0.1}$ \\
$\log\left(\mu_b\right)$ ($\mu_b$ [T])    & $10.2^{+0.1}_{-0.2}$  \\
$\alpha_d$     & $1.9^{+0.9}_{-1.3}$ \\
$\tau_d$ [kyr] & $17.9^{+24.1}_{-14.5}$  \\ 
Birth spacing [yr] & $549^{+292}_{-323}$ \\ 
\hline
\end{tabular}
\end{table}

The contour plot presented in Fig.~\ref{contour_map_ppdot_result} shows that the posterior distributions on the parameters obtained with the SBI pipeline produce magnetars that broadly align with observations. Note that to compare with known magnetars, the only sources included in these contours are those which have undergone an outburst in the last 50 years; we have assumed that all of these would be detected, so we refer to this subpopulation as the ``simulated" group. The known magnetars \citep[from the McGill catalog, ][]{ok14} are shown as red stars in the $P-\dot{P}$ diagram. Furthermore, even magnetars with low $\dot{P}$ are generated and have undergone outbursts in the simulation with a low occurrence rate, matching the observations. When conducting a KS test on the $P$ and $\dot{P}$ distributions of each simulation and the observations, 91 \% of the measured p-values are greater than 0.05 for $\dot{P}$ and 87 \% are greater than 0.05 for $P$, suggesting a significant similarity between these simulations and the observations. This is highlighted in the upper and right panels of Fig.~\ref{contour_map_ppdot_result} where we show the CDFs of the observed population and that of 5000 simulations for $P$ and $\dot{P}$. The CDFs start to diverge from each other slightly at high $P$ and low $\dot{P}$, where we are largely hampered by low number statistics.

\begin{figure}[h]
\centering
\resizebox{\hsize}{!}{\includegraphics{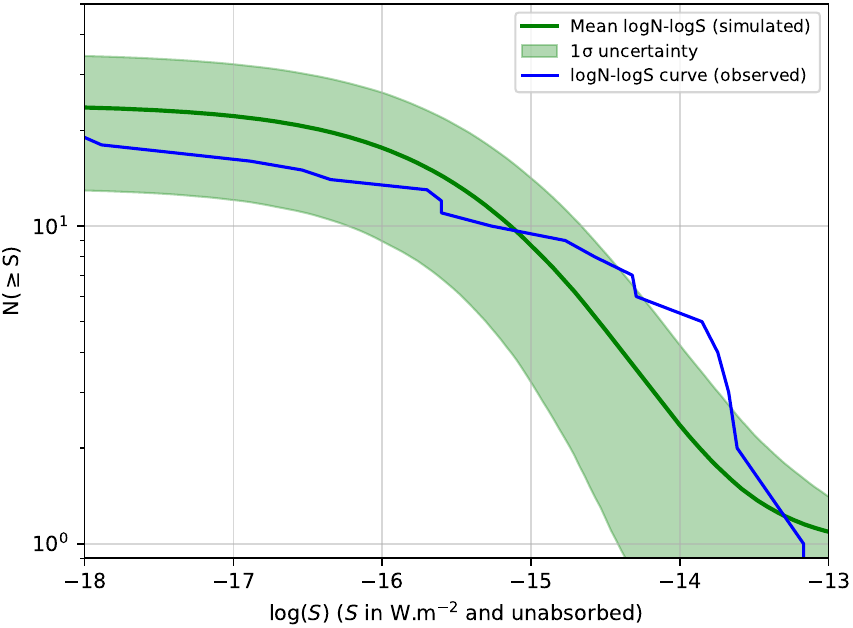}}
\caption{Log$N$-log$S$ plots of the observations compared with the mean log$N$-log$S$ of the mean of 5000 simulations.}
\label{logNlogSresult}
\end{figure}

As another layer of comparison, we generated log$N$-log$S$ distributions for the observed and simulated populations in Fig.\ref{logNlogSresult}. The solid blue line indicates the observed log$N$-log$S$ curve (unabsorbed fluxes were used) and the solid green line indicates the mean log$N$-log$S$ curve of the 5000 post-SBI simulations. The shaded area represents the 1$\sigma$ uncertainty. We can accurately predict the shape of the log$N$-log$S$ within 1~$\sigma$ for most of the flux ranges. There is a small flux range ($10^{-14}$ - $10^{-13.5}$~W.m$^{-2}$) within which the observed values deviate by slightly more than 1~$\sigma$ compared to our simulations.

\begin{figure}[h]
\centering
\resizebox{\hsize}{!}{\includegraphics{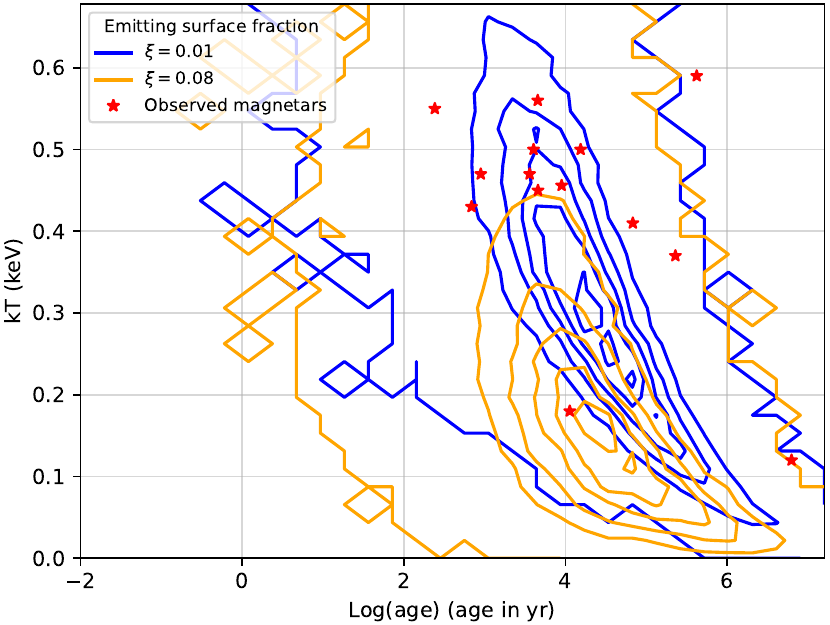}}
\caption{Contour of the temperature versus characteristic age of the simulated magnetars in comparison with the observed magnetars, while choosing $\xi=0.01$ (in blue) and $\xi=0.08$ (in orange) as the emitting surface fraction (it is normalized by the total number of magnetars over 1000 simulations in each bin).}
\label{kt_plots}
\end{figure}

The temperature of the blackbody model for the X-ray emission serves as another point of comparison, as illustrated in Fig.~\ref{kt_plots}. The emitting surface fraction of observed magnetars, defined as $\xi = \left(\frac{R_{eff}}{R}\right)^2$ with $R$ the neutron star radius and $R_{eff} = \sqrt{\frac{d^2 S}{\sigma T^4}}$ (following the blackbody emission definition in equation \eqref{eq:bbem}), yields an average value of $\xi = 0.06$ with a standard deviation of 0.08.
The observed emitting surface fraction rarely exceeds 8\%, which motivated tests with $\xi$ ranging from 1\% to 8\%. However, for clarity, the plots presented here focus only on these two extreme values. The observations indeed fall within the simulated contours in the temperature versus characteristic age plot ($\tau_c = P/2\dot{P}$) when $\xi$ is within this range. These contours suggest that the emitting surface fraction might be relatively low for most magnetars, as lower values of $\xi$ tend to place observed magnetars in more plausible regions of the simulated parameter space. However, the precision of this estimate is primarily constrained by the limited number of known magnetars. Additionally, the lack of detailed knowledge regarding the X-ray emission region of magnetars further hinders the ability to draw robust conclusions on this matter. \\
\begin{figure}[h]
\centering
\resizebox{\hsize}{!}{\includegraphics{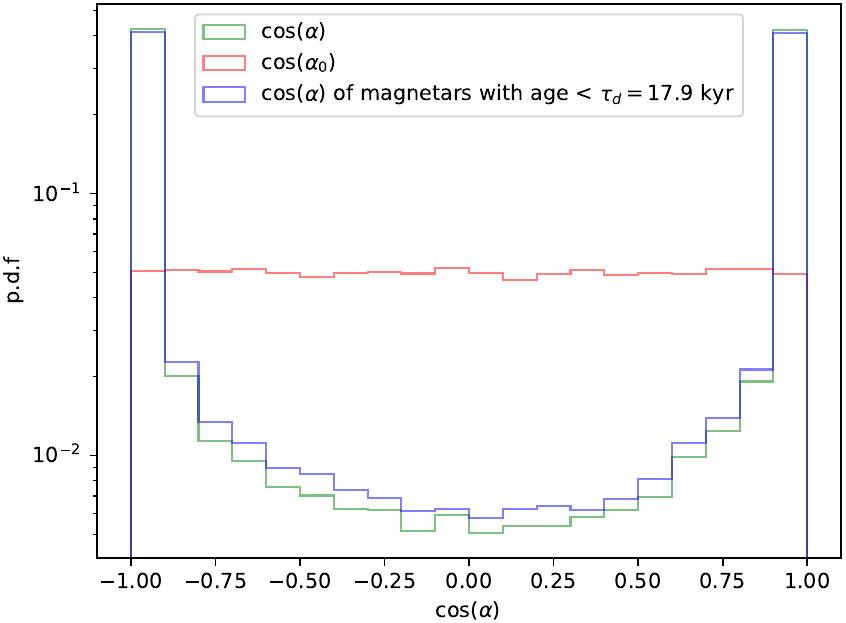}}
\caption{Probability distribution of the inclination angles for simulated magnetars: initial (red), final (green), and those younger than 17.9 kyr (blue), normalized by the total number of magnetars across 5000 simulations in each bin.}
\label{inc_ang_histo}
\end{figure}

The distribution of inclination angles in the simulated population is shown in Fig.~\ref{inc_ang_histo}. Around 80\% of the population exhibits either alignment at 0° or anti-alignment at 180°. The analytical spin-alignment timescale, following \citet{dpm22} and \citet{ptl+14}, is given by:
\begin{equation} \label{eq:spin_align}
\tau_{\rm align}^{\rm MHD} = \frac{I \mu_0 c^3 P_0^2 \sin^2\alpha_0}{16\pi^3 R^6 B_0^2 \cos^4\alpha_0},
\end{equation}
where subscript 0 denotes initial values (except for the vacuum permeability). For a magnetar with $B_0=1.7\times10^{10}$~T (the best estimate of $\mu_b$ with SBI), $P_0=50$~ms, and $\alpha_0=89$°, we obtain $\tau_{\rm align}^{\rm MHD} \approx7\times10^6$\,yr. Conversely, an initial $\alpha_0=91$° results in anti-alignment over the same period. A slight decrease to $\alpha_0=85$° leads to $\tau_{\rm align}^{\rm MHD} \approx11$~kyr, and smaller $\alpha_0$ values further shorten the alignment timescale. Given the best estimate for the magnetic decay timescale $\tau_d=17.9$~kyr, most magnetars align before significant magnetic field decay, as can be seen with the blue distribution representing the magnetars that are younger than $\tau_d=17.9$~kyr. Fig.\ref{inc_ang_histo} also confirms an initially uniform $\alpha_0$ distribution among magnetars, implying rapid alignment regardless of initial inclination. This is consistent with the computed alignment timescales, which become significantly shorter as $\alpha_0$ deviates from 90°. Consequently, only magnetars with $\alpha_0$ near 90° retain a final inclination different from alignment or anti-alignment. As a result, only 20\% of magnetars exhibit non-aligned angles, with fewer than 1.5\% having $\alpha_0$ between 60° and 120°. This behavior contrasts with canonical pulsars, where the  distribution is more uniform and peaks at 90°, making initial inclination a critical factor in their detection (see Fig.~13 of \citealt{dpm22}).

\begin{figure}[h]
\centering
\resizebox{\hsize}{!}{\includegraphics{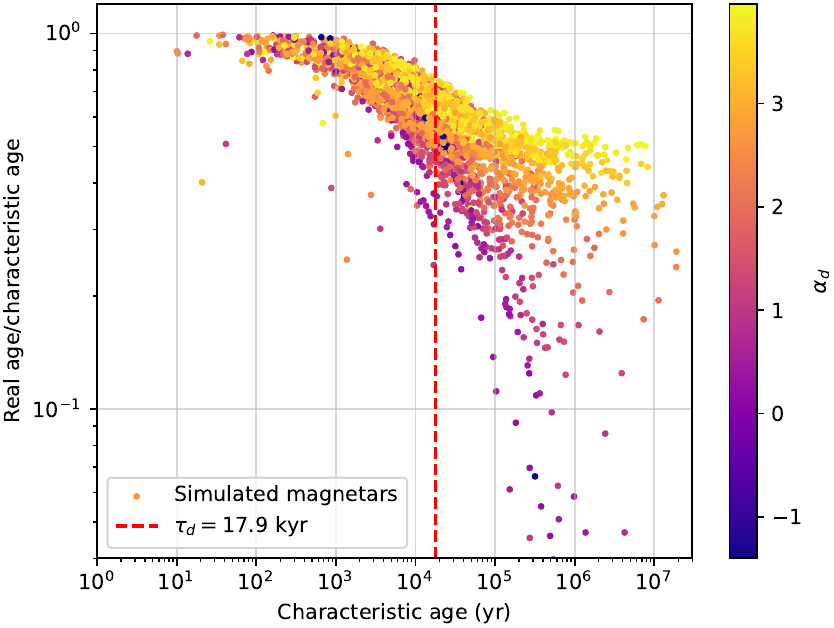}}
\caption{Ratio of real and characteristic age as a function of characteristic age with the parameter $\alpha_d$ used to simulate each magnetar in 120 simulations.}
\label{real_age_charac_age}
\end{figure}

Fig.~\ref{real_age_charac_age} presents the ratio of real to characteristic age for magnetars across 120 simulations, highlighting the impact of $\alpha_d$ on their evolution. Negative $\alpha_d$ leads to a rapid divergence between real and characteristic age once the magnetar surpasses the characteristic magnetic decay timescale. In contrast, for positive $\alpha_d$, the larger the value, the longer it takes for this deviation to occur. The posteriors favor $\alpha_d \approx 2$, implying that only magnetars older than $10^4$~yr—around 20\% of the detected population—experience a significant discrepancy between real and characteristic age. A few young magnetars with $\alpha_d \approx 2$ also show such differences, due to rapid alignment of their rotation and magnetic axes, which quickly results in high $P$ and low $\dot{P}$.

\begin{figure}[h]
\centering
\resizebox{\hsize}{!}{\includegraphics{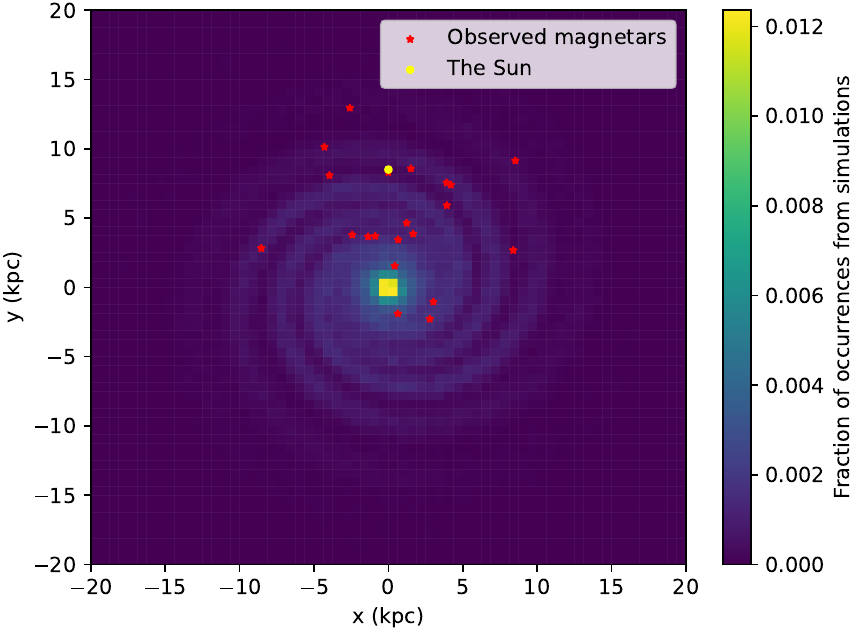}}
\caption{Positions of the simulated magnetars in the Galactic plane (normalized by the total number of magnetars over 5000 simulations in each bin) in comparison with the observed magnetars.}
\label{Positions_in_the_galaxy}
\end{figure}
Recovering the spatial distribution of magnetars presents a challenge in population synthesis. Fig.~\ref{Positions_in_the_galaxy} compares the fraction of magnetars occurring in a given region of the Galactic plane across 5000 simulations with the observed magnetar positions. While the simulated and observed positions do not align, this is expected, as the outburst detection model used here is based solely on energy (which is dependent on magnetic field strength and age) without accounting for distance. The simulations reproduce the shape of the Galactic spiral arms, reflecting the youth of the magnetars which have not migrated significantly from their birth sites. Furthermore, even if distances were incorporated, discrepancies between simulations and observations would persist due to the considerable uncertainties in distance estimates. Including distance in the detection model could nonetheless provide valuable insights into the true spatial distribution of observed magnetars.

\section{Discussion}
\label{sec:discussion}
\subsection{Comparison with other works}

In this study, we estimate the birth rate of magnetars to be ${\rm BR}=1.8^{+2.6}_{-0.6}$~kyr$^{-1}$. The CCSN rate falls between \(11.8 - 21.3 \ \text{kyr}^{-1}\) \citep{rvc21}, within a 68.3\% confidence interval, with approximately 80\% of these CCSNe leading to the formation of isolated neutron stars \citep{fgk06}. Additionally, \citet{spm+24} estimated the birth rate of isolated canonical pulsars at \(24 \ \text{kyr}^{-1}\), placing the overall birth rate of isolated neutron stars between \(9.4 - 24 \ \text{kyr}^{-1}\). Meanwhile, the current star formation rate in the Milky Way is approximately \(1.65 \pm 0.19 \ {\rm M_{\odot} \ yr^{-1}}\) \citep{ln15}. Assuming a Kroupa initial mass function \citep{k01}, this corresponds to the birth rate of stars with \( M > 8 \ {\rm M_{\odot}} \), most of which evolve into neutron stars, yielding an estimated neutron star birth rate of \( 17 \pm 2 \ \text{kyr}^{-1} \) within a 95\% confidence interval.  By comparing the last approximation, which is consistent with the previous estimates, with our magnetar birth rate, we find that $10.7^{+18.7}_{-4.4}$~\% of neutron stars become magnetars. This is close to the 10\% estimate from \citet{ppm10} and provides a well-defined uncertainty on this value, and provides more constrains compared to the estimate of \citet{bhv+19} which reported that more than 12\% of neutron stars become magnetars. This result is on the lower end of the previous estimates, slightly relaxing the birth rate problem of neutron stars \citep{ge24}. 

To obtain a similar simulated population, the magnetic field decay timescale must be quite short; in this study, we find $\tau_d = 17.9^{+24.1}_{-14.5}$~kyr$^{-1}$. This estimation is in agreement with previous studies of the magnetar population that found $\tau_d \sim 10^4$~yr \citep{kds+98,cgp00,dgp12,vrp+13,bhv+19}. 

The mean magnetic field at birth, $\mu_b$, is estimated between $1.15\times10^{10}$~T and $2.40\times10^{10}$~T in our work. This is much greater than the estimate of \citet{gpm+15} and lower than the estimate of \citet{bhv+19}. Interestingly, this range of values is consistent with the magnetic field strengths required for millisecond magnetars to power superluminous supernovae (SLSNe; \citealt{mmk+15,mbg18}) and GRBs \citep{bgm17,mbg18}. The other key factor is the initial spin period; however, the only constraint on its distribution suggests upper limits of 1~s for $\mu_p$ and 0.9 for $\sigma_p$.

\begin{figure*}
    \centering
    \includegraphics[width=\textwidth]{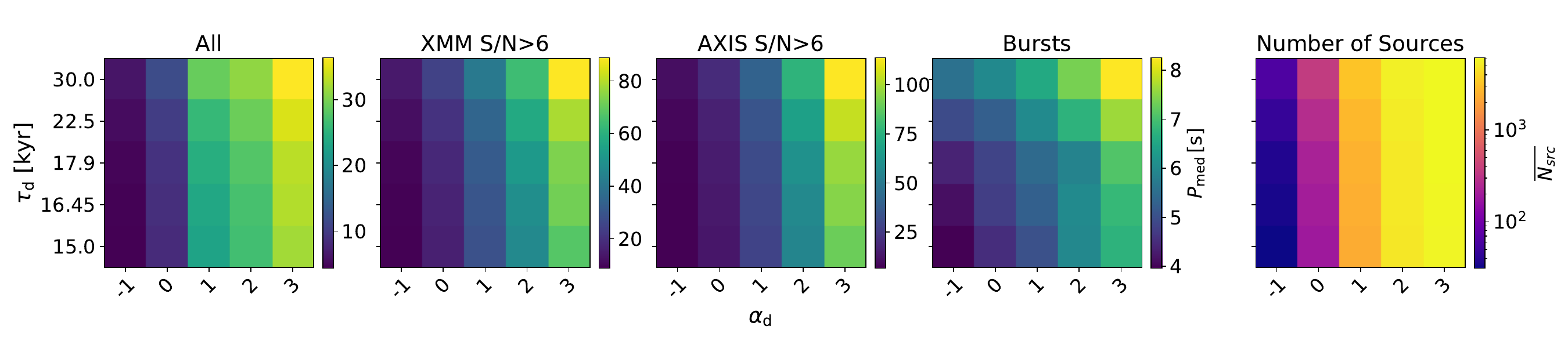}
    \caption{Median Period and Source Counts for Varied Decay Parameters. Holding all other population parameters fixed, $\alpha_{\rm d}$ and $\tau_{\rm d}$ were varied and many populations were produced. The median simulated period was measured from each combination when enough sources had been produced for it to stabilize. In the right-most panel, the average number of sources per population is shown instead. The left three panels are split by detection criteria: the left-most shows all sources, whether or not they are detectable; the second includes sources detected with S/N $>$ 6 in a 10\,ks XMM observation; the third includes those detected with S/N $>$ 6 in a 10\,ks AXIS observation; and the fourth shows sources which have undergone an outburst in the past 50 years.}
    \label{fig:decay_tests}
\end{figure*}

\citet{bhv+19} suggested a relationship between the magnetic field decay parameters and the period distribution of the magnetar population which can be seen in simulations from this study. To demonstrate this, we generated populations of sources while varying $\tau_{\rm d}$ and $\alpha_{\rm d}$ and fixing the remaining population parameters. For each combination of decay parameters, many populations were produced and the median period and total number of sources per population were measured. The number of populations used varied, as increasing alpha dramatically increased the number of magnetars in the population. For larger $\alpha_{\rm d}$, fewer populations could be used without sacrificing the stability in median/source count measurements. Additionally, the populations were filtered by the detection criteria used above to show the difference in detectable and ``full" populations. These results are shown in Figure~\ref{fig:decay_tests}, and show the clear dependence on both maximum period and number of Galactic magnetars.

\subsection{Model improvements and future plans}
While the magnetar evolution model proposed in this work features most of the known aspects of the evolution of these objects, it can be improved. For instance, similar to canonical pulsars, magnetars have been observed to glitch \citep{lmk+09,es10,ybh+23} in many studies. Furthermore, \citet{fer+17} found a relationship for all neutron stars between the glitch activity (a quantity relating the increase in rotation frequency over the time of observations for all the glitches of the concerned neutron star) and the spin-down rate. In the future, we could incorporate glitches in our model with this relationship and test if we are still able to recover a population similar to the observations. Adding this phenomenon would form a more realistic model and may tighten constraints on the parameters which characterize the population of magnetars. Another improvement for the model would be to control the fraction of outburst detected. Outburst events are very energetic, but we are certainly missing some of these (especially less energetic bursts). For example, magnetars XTE~J1810$-$197 and 3XMM J185246.6+003317 were both discovered serendipitously in observations of other targets \citep{ims+04, zcl+14}, with detections showing that these older sources emit weaker bursts with longer tails compared to their younger counterparts. This may suggest that our technique for assigning bursts should weight the burst detectability with the magnetars' ages. Running simulations with a parameter that controls the fraction of detected outbursts (this topic is briefly discussed in \citet{bwt+24}) could help us estimate how many we miss, further improving the accuracy of our results. This parameter would also take the distance into account in the detection, and it would be necessary to compute the energy released during the outburst and its duration. Drawing these two quantities from random distributions (that have to be determined) for each magnetar should work. From this, it would be possible to then compute a flux using the known source position and apply instrument-specific detectability constraints (e.g., Swift-BAT). Finally, another perspective to improve the simulations would be to consider the possible precession of magnetars \citep{m00,apt15,lj18,ip20}. 

\subsection{Detection Prospects}

\begin{figure}
    \centering
    \resizebox{\hsize}{!}{\includegraphics{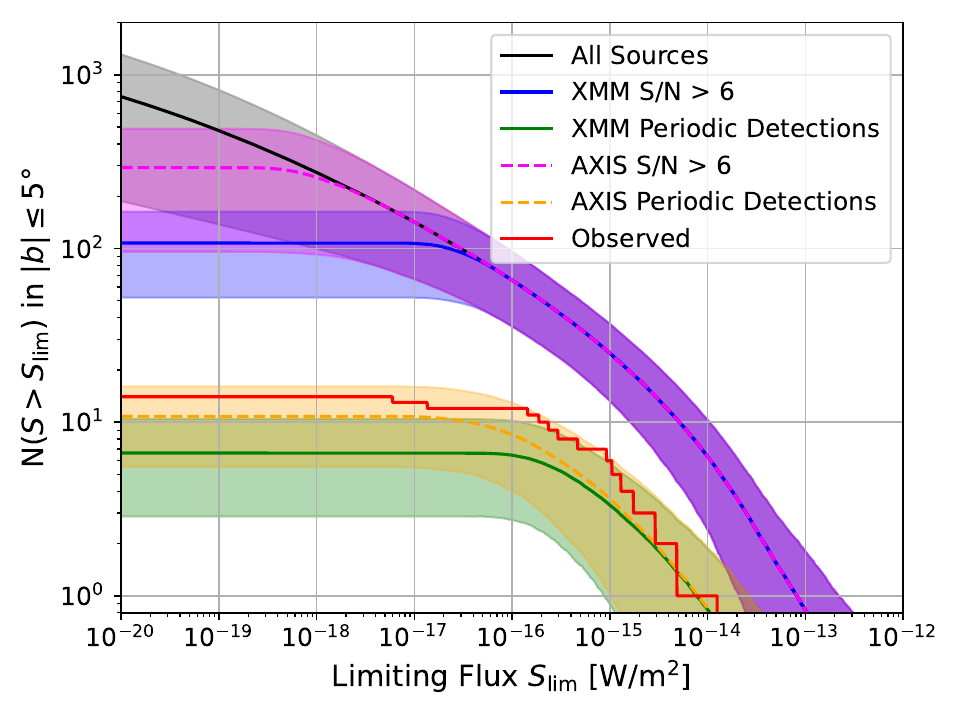}}
    \caption{Detectable magnetar point sources in 1000 simulated populations. The sky region is limited to $|b| \leq 5$, and detection is limited to the $0.4-7\,{\rm keV}$ band. The black band indicates the mean number of sources in all realizations for each limiting flux. In red, known Galactic magnetars are included (only those in the plane with published fluxes are shown). The green band corresponds to sources detected in the periodicity search in XMM data, and blue indicates sources which are detected above 6 S/N in those data. Also included for comparison are comparable results using simulated AXIS response files; these are plotted in orange and magenta. All shaded regions correspond to 1-$\sigma$ uncertainties.}
    \label{fig:detection_lognlogs}
\end{figure}

With a suitable model for the population of magnetars, we can assess the detectability of new magnetar sources. To that end, using realizations of magnetar populations generated with the optimal parameters, we examined all sources above a nominal flux threshold of $10^{-20}$~W\,m$^{-2}$ (i.e, $10^{-17}\,{\rm erg/(s\,cm^2)}$) in the plane ($|b|>5$\,degrees) using the same detection procedure described in Section~\ref{sec:detec}. This formed a mock survey of the sky in the plane, which informs future searching techniques and detection probabilities. For each source, absorption was modeled and a simulated XMM lightcurve was generated for a 10\,ks observation. We applied an H-test and detectable sources were recorded. Here, detectability is broken into three subgroups: the first is simply whether the source has undergone an outburst as predicted using the model described in \S\ref{sec:popevol}; the second is defined as ``point-source" detection, with $\rm S/N > 6$ in a simulated lightcurve; and finally, whether the source is detectable in a periodicity search. Bursting sources are expected to most closely match the true, galactic magnetar population that has been detected. Sources with $\rm S/N > 6$ would not necessarily be identified as magnetars. But periodic detections should include several of the Galactic discoveries as well as a possibly undetected-but-detectable population. This subsample indicates how effective a full-plane search would be, and can be tailored to various instrumentation. For consistency, we simulate XMM observations. Figure~\ref{fig:detection_lognlogs} shows the number of sources above a given limiting flux $S_{\rm lim}$ as a function of $S_{\rm lim}$ for 1000 realizations. This number was chosen based on the stability of the mean value in each bin; above 500 simulations, this mean did not vary substantially, so 1000 ensures an accuracy. This search resulted in 7 periodic detections per simulation on average, and the average number of magnetars above XMM's flux threshold ($\rm \approx10^{-14}\,erg/(s\,cm^2) = 10^{-17}\,W/m^2$) is approximately 150. More sensitive instruments will probe more deeply, and experiments like the Advanced X-ray Imaging Satellite \citep[AXIS; ][]{abc+24} are expected to detect objects fainter by two orders of magnitude \citep{sbb+23}. To test this, we repeated the procedure described above with data simulated using AXIS predictions; the results are shown in Fig.~\ref{fig:detection_lognlogs}. It is apparent that AXIS will detect many more sources, with the potential of doubling periodic detections and more than doubling the number of detectable magnetar point sources.

\begin{figure*}
    \centering
    \includegraphics[width=\linewidth]{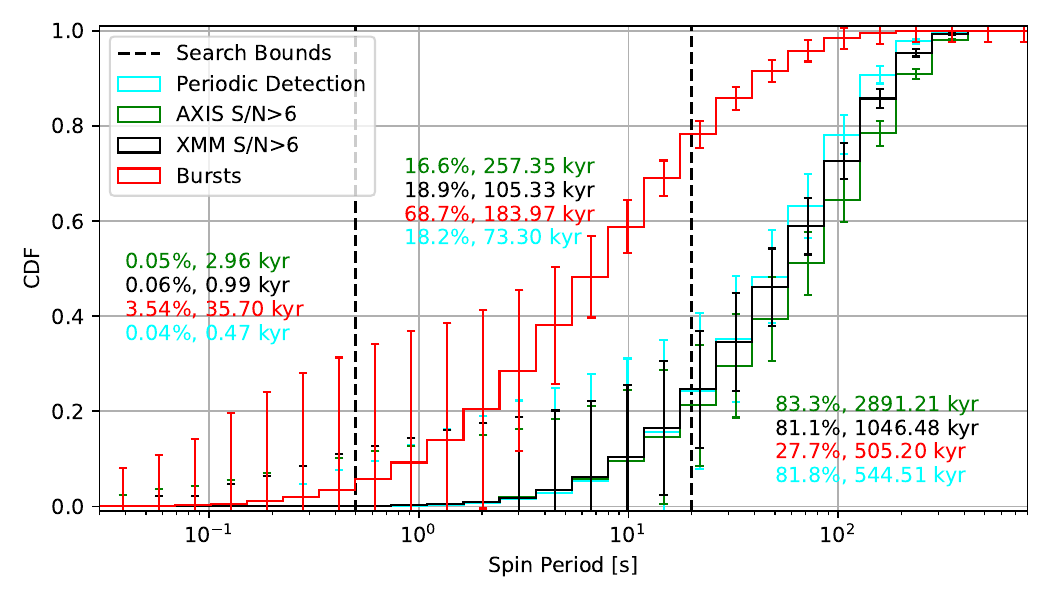}
    \caption{CDFs of simulated sources, split by detection criteria. The dashed lines indicate the period range that was used in the search pipeline, where the left side of the plot indicates periods below 0.5 seconds, those in the center are between 5 and 20 seconds, and the right are above 20 seconds. The green, black, red, and cyan curves are cumulative distributions of the sources from 1000 simulated populations. Error bars indicate $1\sigma$ uncertainties on the distributions. Shown in each period range are the fraction of each detection subpopulation that are within that range and the mean characteristic age of those sources.}
    \label{fig:potdet_cdfs}
\end{figure*}

It is apparent from Figure~\ref{fig:detection_lognlogs} that there is a significant fraction of sources which are detectable as point sources (i.e. $\rm S/N > 6$) but are not detected in the periodicity search, including those on the brightest end of the plot. Examining those sources more closely revealed that the evolution model was producing many sources outside of the periodic search range of $\rm 0.5-20\,s$. Sources in the $\rm 10^{-14}-10^{-13}\,W/m^2$ range correspond to young, bright magnetars with periods below $\rm 0.5\,s$; at the low-flux end, the gap between the two detection populations corresponds to a large number of older magnetars that have evolved to periods longer than $20\,s$. Notably, the former period range includes the periods of both the ``magnetar-like" high-B pulsars J1119$-$6127 (0.4\,s period) and J1846$-$0258 (0.3\,s period). Neither of these sources were initially discovered during outburst \citep{ckl+00,gvb+00}, but both have undergone bursts since \citep{akt+16,bsm+21}. In fact, this model suggests that as many as 80\% of magnetars that could be detected as point sources (and possibly as periodic sources) would effectively be omitted from the search simply because of the search bounds. Figure \ref{fig:potdet_cdfs} shows the CDFs of the spin periods of those sources which are ``detected" according to the three detection criteria. In order to assess determine whether the sources outside of our search range would be detectable in a periodicity search, we have included counts for periodic detections outside of that range. This result is corroborated by the recent discoveries of slowly rotating radio emitters that may be old magnetars \citep{BWM2020,Caleb2022,hzb+22,bwh+23,Wang2024}, and may link magnetars to high-B pulsars \citep{zh00,ckl+00}. Granted, the search region does cover the majority of sources ($\geq 68\%$, assuming all bursts are detected) which have undergone bursting in the last 50 years, reflecting the detection bias toward younger sources with shorter periods.

\section{Conclusions} \label{sec:conclusions}
Simulation-based inference was used in this study to constrain the parameters characterizing the Galactic population of magnetars for the first time. By changing only the parameters for the birth distributions and the magnetic field decay compared to normal pulsars, we used an evolution model that is otherwise identical to normal pulsars and showed it was possible to produce a population of magnetars. This study considered both persistent X-ray detection and outburst detection, taking into account biases from sky coverage and observation information. In the future, with telescopes sensitive enough to detect magnetars through persistent emission alone, this pipeline can be used as-is with only minor changes to the reference population statistics. 

Furthermore, the use of CNNs was essential for extracting features from the density maps, preserving the correlations between the different quantities, and adequately informing the density estimator of SBI. First, the pipeline was validated by a population which was obtained via a known set of parameters. After validation, the following population parameters were constrained: the mean magnetic field at birth $\log\left(\mu_b\right) = 10.2^{+0.1}_{-0.2}$ and standard deviation $\sigma_B = 0.2^{+0.2}_{-0.1}$, the parameter controlling the speed of the magnetic field decay $\alpha_d = 1.9^{+0.9}_{-1.3}$, the typical decay timescale $\tau_d = 17.9^{+24.1}_{-14.5}$~kyr and the birth rate ${\rm BR}=1.8^{+2.6}_{-0.6}$~kyr$^{-1}$. Validation checks showed that the current population is globally insensitive to the initial spin period distribution parameters, $\mu_p$ and $\sigma_p$. However, upper limits of 1~s for $\mu_p$ and 0.9 for $\sigma_p$ seem to be appropriate. The recovered blackbody temperatures of the simulated magnetars suggest that the effective emission radius of magnetars may be quite low for most magnetars, and will require more study to match the temperature versus characteristic age plot with the observations. An analysis of the inclination angle showed that around 80\% of the detected population have aligned (or anti-aligned) rotation and magnetic axes, which implies that seeing pulsation from this population could be quite difficult compared to pulsars.  

Nonetheless, a limitation of the simulation is that the positions of the detected magnetars do not correspond to those of the real magnetars. When accepting a magnetar as detected via outburst, no discrimination through the energy nor the distance is made. Therefore, in order to improve this work in the future, these factors should be taken into account as discussed in the previous section.

Predictions from these results suggest that a full scale Galactic survey for magnetars using XMM would be able to uncover several dozen magnetars with a modest exposure duration. From this relatively shallow mock survey, the population could feasibly double; using upcoming improved instruments like AXIS could potentially quadruple it. Given that many of the results of this study (and other magnetar population studies) are challenged by the small sample size, large-scale searches like those simulated here would be crucially important to fully understanding this population of neutron stars. 

\section*{Acknowledgments}
This work has been supported by the grant ANR-20-CE31-0010.

\software{\texttt{Astropy} \citep{astropy:2018},
\texttt{SciPy} \citep{scipy}, 
\texttt{NumPy} \citep{numpy}, 
\texttt{PyGEDM} \citep{pygedm},
\texttt{PyXSPEC} \citep{pyxspec},
\texttt{SBI} \citep{sbi-soft},
\texttt{TensorFlow} \citep{tensorflow2016}
}

\facilities{XMM-Newton, AXIS}

\phantomsection
\label{sec:AppA}
\section*{Appendix \\ Extended model validation}
\begin{figure}[H]
    \centering
    \resizebox{\hsize}{!}{\includegraphics{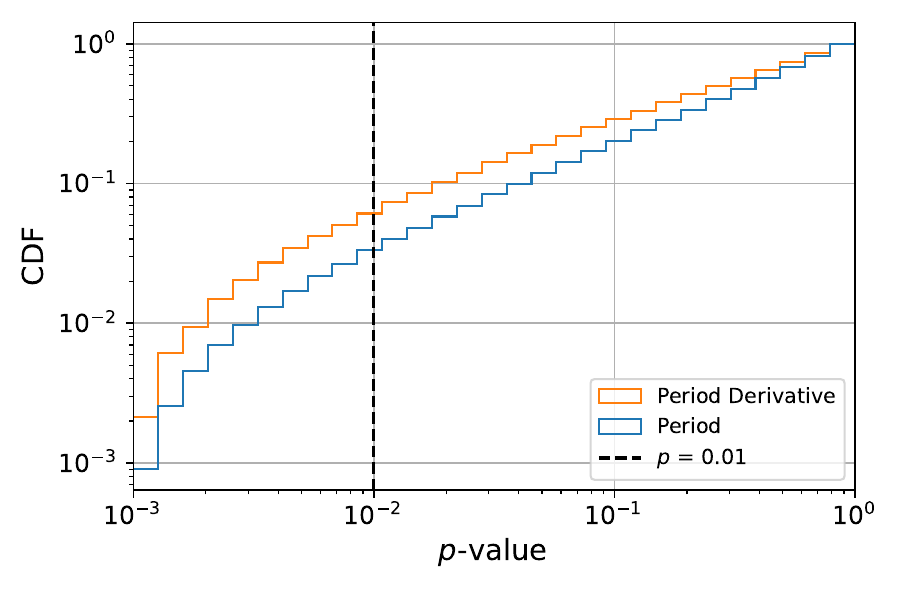}}
    \caption{$p$-values from two-sample KS tests of $P$ and $\dot{P}$ between populations produced using $\vec{\theta_{\rm in}}$ and $\vec{\theta_{\rm out}}$. In total, 20000 population comparisons were made. The vertical dashed line indicates a threshold $p$-value of 0.01; for the two distributions, 2.68\% and 6.08\% lie below this threshold.}
    \label{fig:validation_pvalues}
\end{figure}

To complete the model validation, the Kolmogorov-Smirnov (KS) test \citep{kstest} is applied to the $P$ and $\dot{P}$ distributions between the reference population and each validation simulation. The process was repeated 20000 times, and the distribution of p-values are plotted in Figure~\ref{fig:validation_pvalues}. Those results suggest that in the vast majority of cases the null hypothesis cannot be rejected.

\begin{figure*}
    \centering
    \resizebox{\hsize}{!}{\includegraphics{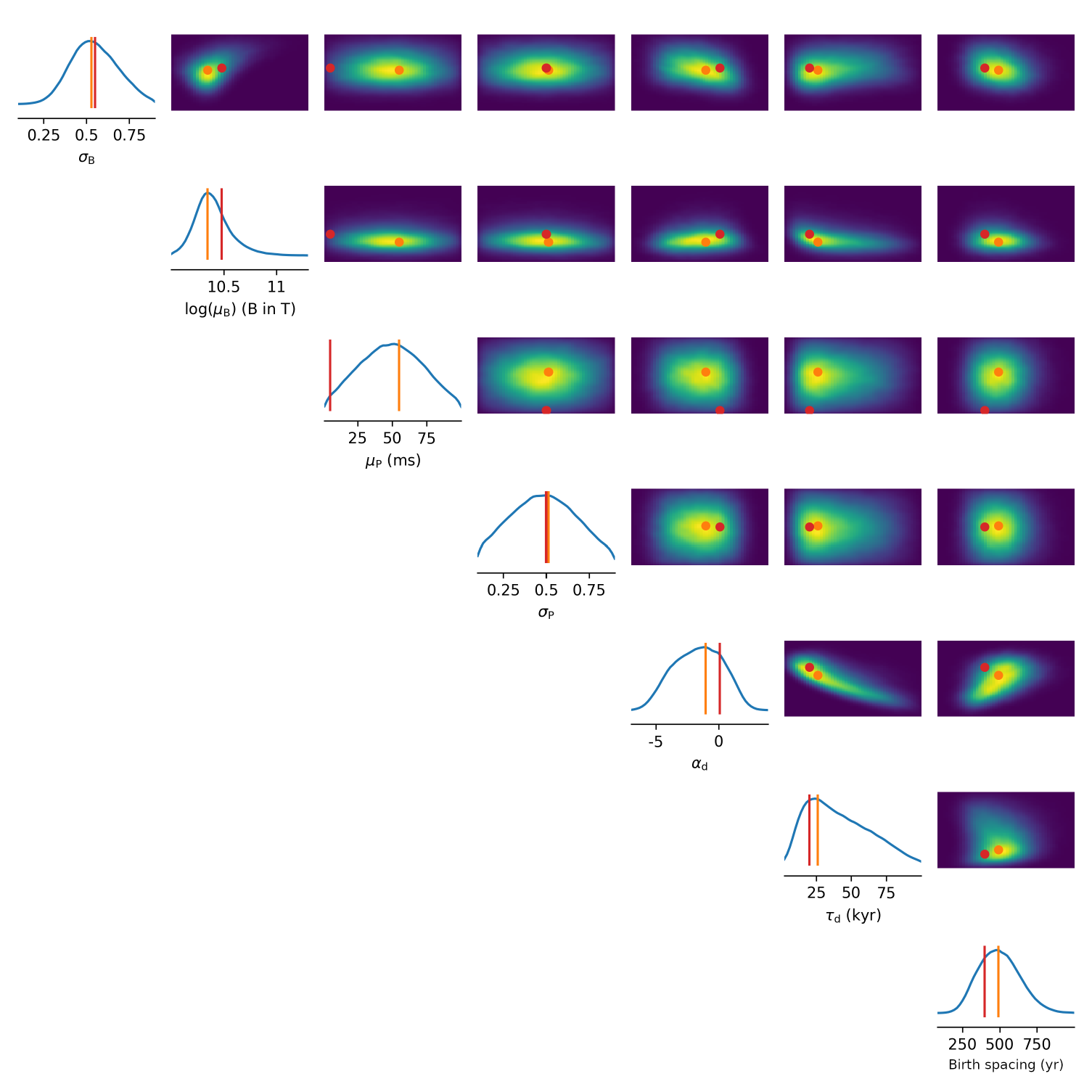}}
    \caption{Corner plot of the posterior distributions of the parameters corresponding to the reference simulation obtained after one validation run. Posterior medians are in yellow and the ground truths in red.}
    \label{cornerplot_valid}
\end{figure*}

The posterior distributions obtained in Fig.\ref{cornerplot_valid} are an example of the results from one of the validation runs mentioned in \S\ref{sec:validation}. In this figure, it is apparent that in each distribution the ground truth (in red, representing $\vec{\theta}_{\rm in}$) is close to the median of the posterior found (shown in yellow, $\vec{\theta}_{\rm out}$), except for the distribution of $\mu_p$. Even though in this run $\sigma_p$ is well recovered, the distribution for this parameter is quite spread, indicating that it also seems insensitive to this parameter as evidenced by Fig.~\ref{fig:validation}.

\clearpage
\bibliographystyle{apj}
\bibliography{xmm}

\end{document}

%% file: tables/constants.tex
\begin{table}[h]
\caption{Constant values used for the different potentials, with the solar mass $M_{\odot}=1.99 \times 10^{30}$ kg.} 
\label{tabl_const_pot} 
\centering 
\begin{tabular}{c c} 
\hline\hline 
Parameters & Values \\
\hline 
$M_h$ & $2.9 \times 10^{11} \pm 7.6 \times 10^{10} M_{\odot}$  \\ 
$a_h$ & 7.7 $\pm$ 2.1 kpc \\
$a_d$ & 4.4 $\pm$ 0.73 kpc \\
$b_d$ & 0.308 $\pm$ 0.005 kpc  \\
$M_d$ & $6.50 \times 10^{10} \pm 1.9 \times 10^9  M_{\odot}$ \\
$a_b$ & 0.0 kpc \\
$b_b$ & 0.267 $\pm$ 0.009 kpc \\
$M_b$ & $1.02 \times 10^{10} \pm 6.3 \times 10^8 M_{\odot}$ \\
$M_{n}$ & $4.10^6 \pm 0.42 \times 10^6 M_{\odot}$ \\
\hline 
\end{tabular}
\end{table} 

%% file: tables/input_parameter_ranges.tex
\begin{table*}
    \centering
    \begin{tabular}{l|c|c|c}
    \hline\hline 
    Quantity & Parameter & Range & Unit \\
    \hline 
    Standard deviation of birth magnetic field & $\sigma_b$ & [0.1,0.9] \\
    Standard deviation of birth period& $\sigma_p$ & [0.1,0.9] \\
    Mean birth period & $\mu_p$ & [1,100] & ms \\
    Mean birth magnetic field & $\mu_b$ & [$1\times10^{10}$,$2\times10^{11}$] & T \\
    Field decay scaling & $\alpha_d$ & [-7,4] &  \\
    Field decay timescale & $\tau_d$ & [2,100] & kyr \\ 
    Birth rate & $BR$ & [1/1000,1/50] & yr$^{-1}$ \\ 
    \end{tabular}
    \caption{Ranges of input parameters for SBI procedure.}
    \label{tab:inparam}
\end{table*}

%% file: main.bbl
\begin{thebibliography}{136}
\expandafter\ifx\csname natexlab\endcsname\relax\def\natexlab#1{#1}\fi

\bibitem[{Abadi {et~al.}(2016)Abadi, Agarwal, {et~al.}}]{tensorflow2016}
Abadi, M., Agarwal, A., {et~al.} 2016, TensorFlow: Large-Scale Machine Learning
  on Heterogeneous Distributed Systems

\bibitem[{{Alsing} {et~al.}(2019){Alsing}, {Charnock}, {Feeney}, \&
  {Wandelt}}]{acf+19}
{Alsing}, J., {Charnock}, T., {Feeney}, S., \& {Wandelt}, B. 2019, \mnras, 488,
  4440

\bibitem[{{Archibald} {et~al.}(2016){Archibald}, {Kaspi}, {Tendulkar}, \&
  {Scholz}}]{akt+16}
{Archibald}, R.~F., {Kaspi}, V.~M., {Tendulkar}, S.~P., \& {Scholz}, P. 2016,
  \apjl, 829, L21

\bibitem[{{Arcodia} {et~al.}(2024){Arcodia}, {Bauer}, {Cenko}, {Dage},
  {Haggard}, {Ho}, {Kara}, {Koss}, {Liu}, {Mallick}, {Negro}, {Pradhan},
  {Quirola-V{\'a}squez}, {Reynolds}, {Ricci}, {Rothschild}, {Sridhar}, {Troja},
  \& {Yao}}]{abc+24}
{Arcodia}, R., {et~al.} 2024, Universe, 10, 316

\bibitem[{{Arnaud}(1996)}]{a96}
{Arnaud}, K.~A. 1996, in Astronomical Society of the Pacific Conference Series,
  Vol. 101, Astronomical Data Analysis Software and Systems V, ed. G.~H.
  {Jacoby} \& J.~{Barnes}, 17

\bibitem[{{Arzamasskiy} {et~al.}(2015){Arzamasskiy}, {Philippov}, \&
  {Tchekhovskoy}}]{apt15}
{Arzamasskiy}, L., {Philippov}, A., \& {Tchekhovskoy}, A. 2015, \mnras, 453,
  3540

\bibitem[{{Bajkova} \& {Bobylev}(2021)}]{bb+21}
{Bajkova}, A.~T., \& {Bobylev}, V.~V. 2021, Astronomical and Astrophysical
  Transactions, 32, 177

\bibitem[{{Barr{\`e}re} {et~al.}(2024){Barr{\`e}re}, {Guilet}, {Raynaud}, \&
  {Reboul-Salze}}]{bgr+24}
{Barr{\`e}re}, P., {Guilet}, J., {Raynaud}, R., \& {Reboul-Salze}, A. 2024,
  arXiv e-prints, arXiv:2407.01775

\bibitem[{{Beniamini} {et~al.}(2017){Beniamini}, {Giannios}, \&
  {Metzger}}]{bgm17}
{Beniamini}, P., {Giannios}, D., \& {Metzger}, B.~D. 2017, \mnras, 472, 3058

\bibitem[{{Beniamini} {et~al.}(2019){Beniamini}, {Hotokezaka}, {van der Horst},
  \& {Kouveliotou}}]{bhv+19}
{Beniamini}, P., {Hotokezaka}, K., {van der Horst}, A., \& {Kouveliotou}, C.
  2019, \mnras, 487, 1426

\bibitem[{{Beniamini} \& {Kumar}(2023)}]{BK2023}
{Beniamini}, P., \& {Kumar}, P. 2023, \mnras, 519, 5345

\bibitem[{{Beniamini} \& {Kumar}(2024)}]{BK2024}
---. 2024, arXiv e-prints, arXiv:2410.19043

\bibitem[{{Beniamini} \& {Lu}(2021)}]{bl21}
{Beniamini}, P., \& {Lu}, W. 2021, \apj, 920, 109

\bibitem[{{Beniamini} \& {Piran}(2024)}]{bp24}
{Beniamini}, P., \& {Piran}, T. 2024, \apj, 966, 17

\bibitem[{{Beniamini} {et~al.}(2023){Beniamini}, {Wadiasingh}, {Hare},
  {Rajwade}, {Younes}, \& {van der Horst}}]{bwh+23}
{Beniamini}, P., {Wadiasingh}, Z., {Hare}, J., {Rajwade}, K.~M., {Younes}, G.,
  \& {van der Horst}, A.~J. 2023, \mnras, 520, 1872

\bibitem[{{Beniamini} {et~al.}(2020){Beniamini}, {Wadiasingh}, \&
  {Metzger}}]{BWM2020}
{Beniamini}, P., {Wadiasingh}, Z., \& {Metzger}, B.~D. 2020, \mnras, 496, 3390

\bibitem[{{Beniamini} {et~al.}(2024){Beniamini}, {Wadiasingh}, {Trigg},
  {Chirenti}, {Burns}, {Younes}, {Negro}, \& {Granot}}]{bwt+24}
{Beniamini}, P., {Wadiasingh}, Z., {Trigg}, A., {Chirenti}, C., {Burns}, E.,
  {Younes}, G., {Negro}, M., \& {Granot}, J. 2024, arXiv e-prints,
  arXiv:2411.16846

\bibitem[{{Berteaud} {et~al.}(2024){Berteaud}, {Eckner}, {Calore}, {Clavel}, \&
  {Haggard}}]{bec+24}
{Berteaud}, J., {Eckner}, C., {Calore}, F., {Clavel}, M., \& {Haggard}, D.
  2024, \apj, 974, 144

\bibitem[{{Beskin} {et~al.}(1993){Beskin}, {Gurevich}, \& {Istomin}}]{bgi93}
{Beskin}, V.~S., {Gurevich}, A.~V., \& {Istomin}, Y.~N. 1993, {Physics of the
  pulsar magnetosphere}

\bibitem[{{Beskin} {et~al.}(2013){Beskin}, {Istomin}, \& {Philippov}}]{bip13}
{Beskin}, V.~S., {Istomin}, Y.~N., \& {Philippov}, A.~A. 2013, Physics Uspekhi,
  56, 164

\bibitem[{{Beskin} \& {Nokhrina}(2007)}]{bn07}
{Beskin}, V.~S., \& {Nokhrina}, E.~E. 2007, \apss, 308, 569

\bibitem[{{Bhardwaj} {et~al.}(2023){Bhardwaj}, {Alvey}, {Miller}, {Nissanke},
  \& {Weniger}}]{bam+23}
{Bhardwaj}, U., {Alvey}, J., {Miller}, B.~K., {Nissanke}, S., \& {Weniger}, C.
  2023, \prd, 108, 042004

\bibitem[{{Bhattacharya} {et~al.}(1992){Bhattacharya}, {Wijers}, {Hartman}, \&
  {Verbunt}}]{bwh+92}
{Bhattacharya}, D., {Wijers}, R. A.~M.~J., {Hartman}, J.~W., \& {Verbunt}, F.
  1992, \aap, 254, 198

\bibitem[{{Blumer} {et~al.}(2021){Blumer}, {Safi-Harb}, {McLaughlin}, \&
  {Fiore}}]{bsm+21}
{Blumer}, H., {Safi-Harb}, S., {McLaughlin}, M.~A., \& {Fiore}, W. 2021, \apjl,
  911, L6

\bibitem[{{Bochenek} {et~al.}(2020){Bochenek}, {Ravi}, {Belov}, {Hallinan},
  {Kocz}, {Kulkarni}, \& {McKenna}}]{brb+20}
{Bochenek}, C.~D., {Ravi}, V., {Belov}, K.~V., {Hallinan}, G., {Kocz}, J.,
  {Kulkarni}, S.~R., \& {McKenna}, D.~L. 2020, \nat, 587, 59

\bibitem[{{Bogdanov} {et~al.}(2019){Bogdanov}, {Lamb}, {Mahmoodifar}, {Miller},
  {Morsink}, {Riley}, {Strohmayer}, {Tung}, {Watts}, {Dittmann}, {Chakrabarty},
  {Guillot}, {Arzoumanian}, \& {Gendreau}}]{blm+19}
{Bogdanov}, S., {et~al.} 2019, \apjl, 887, L26

\bibitem[{{Bovy}(2015)}]{b+15}
{Bovy}, J. 2015, \apjs, 216, 29

\bibitem[{{Caleb} {et~al.}(2022){Caleb}, {Heywood}, {Rajwade}, {Malenta},
  {Stappers}, {Barr}, {Chen}, {Morello}, {Sanidas}, {van den Eijnden},
  {Kramer}, {Buckley}, {Brink}, {Motta}, {Woudt}, {Weltevrede}, {Jankowski},
  {Surnis}, {Buchner}, {Bezuidenhout}, {Driessen}, \& {Fender}}]{Caleb2022}
{Caleb}, M., {et~al.} 2022, Nature Astronomy, 6, 828

\bibitem[{{Camilo} {et~al.}(2000){Camilo}, {Kaspi}, {Lyne}, {Manchester},
  {Bell}, {D'Amico}, {McKay}, \& {Crawford}}]{ckl+00}
{Camilo}, F., {Kaspi}, V.~M., {Lyne}, A.~G., {Manchester}, R.~N., {Bell},
  J.~F., {D'Amico}, N., {McKay}, N.~P.~F., \& {Crawford}, F. 2000, \apj, 541,
  367

\bibitem[{Collazzi {et~al.}(2015)Collazzi, Kouveliotou, Horst, Younes, Kaneko,
  Göğüş, Lin, Granot, Finger, Chaplin, Huppenkothen, Watts, Kienlin,
  Baring, Gruber, Bhat, Gibby, Gehrels, McEnery, Klis, \& Wijers}]{collazzi}
Collazzi, A.~C., {et~al.} 2015, The Astrophysical Journal Supplement Series,
  218, 11

\bibitem[{{Colpi} {et~al.}(2000){Colpi}, {Geppert}, \& {Page}}]{cgp00}
{Colpi}, M., {Geppert}, U., \& {Page}, D. 2000, \apjl, 529, L29

\bibitem[{{Cordes} \& {Chatterjee}(2019)}]{cc19}
{Cordes}, J.~M., \& {Chatterjee}, S. 2019, \araa, 57, 417

\bibitem[{{Coti Zelati} {et~al.}(2018){Coti Zelati}, {Rea}, {Pons}, {Campana},
  \& {Esposito}}]{cotizelati}
{Coti Zelati}, F., {Rea}, N., {Pons}, J.~A., {Campana}, S., \& {Esposito}, P.
  2018, \mnras, 474, 961

\bibitem[{{Cowsik}(1998)}]{c98}
{Cowsik}, R. 1998, \aap, 340, L65

\bibitem[{{Cranmer} {et~al.}(2020){Cranmer}, {Brehmer}, \& {Louppe}}]{cbl+20}
{Cranmer}, K., {Brehmer}, J., \& {Louppe}, G. 2020, Proceedings of the National
  Academy of Science, 117, 30055

\bibitem[{{Cumming} {et~al.}(2004){Cumming}, {Arras}, \& {Zweibel}}]{caa+04}
{Cumming}, A., {Arras}, P., \& {Zweibel}, E. 2004, \apj, 609, 999

\bibitem[{{Dai} \& {Lu}(1998)}]{dl98}
{Dai}, Z.~G., \& {Lu}, T. 1998, \aap, 333, L87

\bibitem[{{Dall'Osso} {et~al.}(2012){Dall'Osso}, {Granot}, \& {Piran}}]{dgp12}
{Dall'Osso}, S., {Granot}, J., \& {Piran}, T. 2012, \mnras, 422, 2878

\bibitem[{{de Jager} \& {B{\"u}sching}(2010)}]{db10}
{de Jager}, O.~C., \& {B{\"u}sching}, I. 2010, \aap, 517, L9

\bibitem[{{Dirson} {et~al.}(2022){Dirson}, {P{\'e}tri}, \& {Mitra}}]{dpm22}
{Dirson}, L., {P{\'e}tri}, J., \& {Mitra}, D. 2022, \aap, 667, A82

\bibitem[{{Doroshenko}(2024)}]{d24}
{Doroshenko}, V. 2024, arXiv e-prints, arXiv:2403.03127

\bibitem[{{Duncan} \& {Thompson}(1992)}]{dt+92}
{Duncan}, R.~C., \& {Thompson}, C. 1992, \apjl, 392, L9

\bibitem[{{Eichler} \& {Shaisultanov}(2010)}]{es10}
{Eichler}, D., \& {Shaisultanov}, R. 2010, \apjl, 715, L142

\bibitem[{{Faucher-Gigu{\`e}re} \& {Kaspi}(2006)}]{fgk06}
{Faucher-Gigu{\`e}re}, C.-A., \& {Kaspi}, V.~M. 2006, \apj, 643, 332

\bibitem[{{Foglizzo} {et~al.}(2015){Foglizzo}, {Kazeroni}, {Guilet}, {Masset},
  {Gonz{\'a}lez}, {Krueger}, {Novak}, {Oertel}, {Margueron}, {Faure}, {Martin},
  {Blottiau}, {Peres}, \& {Durand}}]{fkg+15}
{Foglizzo}, T., {et~al.} 2015, \pasa, 32, e009

\bibitem[{{Fuentes} {et~al.}(2017){Fuentes}, {Espinoza}, {Reisenegger}, {Shaw},
  {Stappers}, \& {Lyne}}]{fer+17}
{Fuentes}, J.~R., {Espinoza}, C.~M., {Reisenegger}, A., {Shaw}, B., {Stappers},
  B.~W., \& {Lyne}, A.~G. 2017, \aap, 608, A131

\bibitem[{{Gen{\c{c}}ali} \& {Ertan}(2024)}]{ge24}
{Gen{\c{c}}ali}, A.~A., \& {Ertan}, {\"U}. 2024, \mnras, 534, 1481

\bibitem[{{Gordon} {et~al.}(2023){Gordon}, {Fong}, {Kilpatrick}, {Eftekhari},
  {Leja}, {Prochaska}, {Nugent}, {Bhandari}, {Blanchard}, {Caleb}, {Day},
  {Deller}, {Dong}, {Glowacki}, {Gourdji}, {Mannings}, {Mahoney}, {Marnoch},
  {Miller}, {Paterson}, {Rastinejad}, {Ryder}, {Sadler}, {Scott}, {Sears},
  {Shannon}, {Simha}, {Stappers}, \& {Tejos}}]{Gordon2023}
{Gordon}, A.~C., {et~al.} 2023, arXiv e-prints, arXiv:2302.05465

\bibitem[{{Gordon} \& {Arnaud}(2021)}]{pyxspec}
{Gordon}, C., \& {Arnaud}, K. 2021, {PyXspec: Python interface to XSPEC
  spectral-fitting program}, Astrophysics Source Code Library, record
  ascl:2101.014

\bibitem[{{Gotthelf} {et~al.}(2000){Gotthelf}, {Vasisht}, {Boylan-Kolchin}, \&
  {Torii}}]{gvb+00}
{Gotthelf}, E.~V., {Vasisht}, G., {Boylan-Kolchin}, M., \& {Torii}, K. 2000,
  \apjl, 542, L37

\bibitem[{{G{\"o}{\v{g}}{\"u}{\c{S}} } {et~al.}(1999){G{\"o}{\v{g}}{\"u}{\c{S}}
  }, {Woods}, {Kouveliotou}, {van Paradijs}, {Briggs}, {Duncan}, \&
  {Thompson}}]{gwk+99}
{G{\"o}{\v{g}}{\"u}{\c{S}} }, E., {Woods}, P.~M., {Kouveliotou}, C., {van
  Paradijs}, J., {Briggs}, M.~S., {Duncan}, R.~C., \& {Thompson}, C. 1999,
  \apjl, 526, L93

\bibitem[{{G{\"o}{\v{g}}{\"u}{\c{s}}}
  {et~al.}(2000){G{\"o}{\v{g}}{\"u}{\c{s}}}, {Woods}, {Kouveliotou}, {van
  Paradijs}, {Briggs}, {Duncan}, \& {Thompson}}]{gwk+00}
{G{\"o}{\v{g}}{\"u}{\c{s}}}, E., {Woods}, P.~M., {Kouveliotou}, C., {van
  Paradijs}, J., {Briggs}, M.~S., {Duncan}, R.~C., \& {Thompson}, C. 2000,
  \apjl, 532, L121

\bibitem[{{Graber} {et~al.}(2024){Graber}, {Ronchi}, {Pardo-Araujo}, \&
  {Rea}}]{grp+24}
{Graber}, V., {Ronchi}, M., {Pardo-Araujo}, C., \& {Rea}, N. 2024, \apj, 968,
  16

\bibitem[{{Gull{\'o}n} {et~al.}(2015){Gull{\'o}n}, {Pons}, {Miralles},
  {Vigan{\`o}}, {Rea}, \& {Perna}}]{gpm+15}
{Gull{\'o}n}, M., {Pons}, J.~A., {Miralles}, J.~A., {Vigan{\`o}}, D., {Rea},
  N., \& {Perna}, R. 2015, \mnras, 454, 615

\bibitem[{{Han} {et~al.}(2021){Han}, {Wang}, {Wang}, {Wang}, {Zhou}, {Sun},
  {Yan}, {Su}, {Jing}, {Chen}, {Gao}, {Hou}, {Xu}, {Lee}, {Wang}, {Jiang},
  {Xu}, {Yan}, {Gan}, {Guan}, {Huang}, {Jiang}, {Li}, {Men}, {Sun}, {Wang},
  {Wang}, {Wang}, {Xie}, {Xu}, {Yao}, {You}, {Yu}, {Yuan}, {Yuen}, {Zhang}, \&
  {Zhu}}]{hww+21}
{Han}, J.~L., {et~al.} 2021, Research in Astronomy and Astrophysics, 21, 107

\bibitem[{{Hermans} {et~al.}(2019){Hermans}, {Begy}, \& {Louppe}}]{hbl+19}
{Hermans}, J., {Begy}, V., \& {Louppe}, G. 2019, arXiv e-prints,
  arXiv:1903.04057

\bibitem[{{Hobbs} {et~al.}(2005){Hobbs}, {Lorimer}, {Lyne}, \&
  {Kramer}}]{hll+05}
{Hobbs}, G., {Lorimer}, D.~R., {Lyne}, A.~G., \& {Kramer}, M. 2005, \mnras,
  360, 974

\bibitem[{{Hu} {et~al.}(2019){Hu}, {Ng}, \& {Ho}}]{hnh19}
{Hu}, C.-P., {Ng}, C.~Y., \& {Ho}, W. C.~G. 2019, \mnras, 485, 4274

\bibitem[{{Hurley-Walker} {et~al.}(2022){Hurley-Walker}, {Zhang}, {Bahramian},
  {McSweeney}, {O'Doherty}, {Hancock}, {Morgan}, {Anderson}, {Heald}, \&
  {Galvin}}]{hzb+22}
{Hurley-Walker}, N., {et~al.} 2022, \nat, 601, 526

\bibitem[{{Ibrahim} {et~al.}(2004){Ibrahim}, {Markwardt}, {Swank}, {Ransom},
  {Roberts}, {Kaspi}, {Woods}, {Safi-Harb}, {Balman}, {Parke}, {Kouveliotou},
  {Hurley}, \& {Cline}}]{ims+04}
{Ibrahim}, A.~I., {et~al.} 2004, \apjl, 609, L21

\bibitem[{{Igoshev} {et~al.}(2022){Igoshev}, {Frantsuzova}, {Gourgouliatos},
  {Tsichli}, {Konstantinou}, \& {Popov}}]{ifg+22}
{Igoshev}, A.~P., {Frantsuzova}, A., {Gourgouliatos}, K.~N., {Tsichli}, S.,
  {Konstantinou}, L., \& {Popov}, S.~B. 2022, \mnras, 514, 4606

\bibitem[{{Igoshev} \& {Popov}(2020)}]{ip20}
{Igoshev}, A.~P., \& {Popov}, S.~B. 2020, \mnras, 499, 2826

\bibitem[{{Igoshev} \& {Popov}(2024)}]{ip24}
---. 2024, \mnras, 535, L54

\bibitem[{{Jawor} \& {Tauris}(2022)}]{jt+22}
{Jawor}, J.~A., \& {Tauris}, T.~M. 2022, \mnras, 509, 634

\bibitem[{Jones {et~al.}(2001)Jones, Oliphant, Peterson, {et~al.}}]{scipy}
Jones, E., Oliphant, T., Peterson, P., {et~al.} 2001, {SciPy}: Open source
  scientific tools for {Python}

\bibitem[{{Kapil} {et~al.}(2023){Kapil}, {Mandel}, {Berti}, \&
  {M{\"u}ller}}]{kmb+23}
{Kapil}, V., {Mandel}, I., {Berti}, E., \& {M{\"u}ller}, B. 2023, \mnras, 519,
  5893

\bibitem[{{Kasen} \& {Bildsten}(2010)}]{kb10}
{Kasen}, D., \& {Bildsten}, L. 2010, \apj, 717, 245

\bibitem[{{Kaspi} \& {Beloborodov}(2017)}]{kb+17}
{Kaspi}, V.~M., \& {Beloborodov}, A.~M. 2017, \araa, 55, 261

\bibitem[{{Kouveliotou} {et~al.}(1994){Kouveliotou}, {Fishman}, {Meegan},
  {Paciesas}, {van Paradijs}, {Norris}, {Preece}, {Briggs}, {Horack},
  {Pendleton}, \& {Green}}]{kfm+94}
{Kouveliotou}, C., {et~al.} 1994, \nat, 368, 125

\bibitem[{{Kouveliotou} {et~al.}(1998){Kouveliotou}, {Dieters}, {Strohmayer},
  {van Paradijs}, {Fishman}, {Meegan}, {Hurley}, {Kommers}, {Smith}, {Frail},
  \& {Murakami}}]{kds+98}
---. 1998, \nat, 393, 235

\bibitem[{{Kroupa}(2001)}]{k01}
{Kroupa}, P. 2001, \mnras, 322, 231

\bibitem[{Kullback \& Leibler(1951)}]{kl+51}
Kullback, S., \& Leibler, R.~A. 1951, The Annals of Mathematical Statistics,
  22, 79

\bibitem[{{Kumar} {et~al.}(2017){Kumar}, {Lu}, \& {Bhattacharya}}]{kumar+17}
{Kumar}, P., {Lu}, W., \& {Bhattacharya}, M. 2017, \mnras, 468, 2726

\bibitem[{{Lander} \& {Jones}(2018)}]{lj18}
{Lander}, S.~K., \& {Jones}, D.~I. 2018, \mnras, 481, 4169

\bibitem[{{Licquia} \& {Newman}(2015)}]{ln15}
{Licquia}, T.~C., \& {Newman}, J.~A. 2015, \apj, 806, 96

\bibitem[{{Lu} {et~al.}(2020){Lu}, {Kumar}, \& {Zhang}}]{lkz20}
{Lu}, W., {Kumar}, P., \& {Zhang}, B. 2020, \mnras, 498, 1397

\bibitem[{{Lyne} {et~al.}(2015){Lyne}, {Jordan}, {Graham-Smith}, {Espinoza},
  {Stappers}, \& {Weltevrede}}]{ljg+15}
{Lyne}, A.~G., {Jordan}, C.~A., {Graham-Smith}, F., {Espinoza}, C.~M.,
  {Stappers}, B.~W., \& {Weltevrede}, P. 2015, \mnras, 446, 857

\bibitem[{{Lyne} {et~al.}(2009){Lyne}, {McLaughlin}, {Keane}, {Kramer},
  {Espinoza}, {Stappers}, {Palliyaguru}, \& {Miller}}]{lmk+09}
{Lyne}, A.~G., {McLaughlin}, M.~A., {Keane}, E.~F., {Kramer}, M., {Espinoza},
  C.~M., {Stappers}, B.~W., {Palliyaguru}, N.~T., \& {Miller}, J. 2009, \mnras,
  400, 1439

\bibitem[{{Manchester} {et~al.}(2001){Manchester}, {Lyne}, {Camilo}, {Bell},
  {Kaspi}, {D'Amico}, {McKay}, {Crawford}, {Stairs}, {Possenti}, {Kramer}, \&
  {Sheppard}}]{mlc+01}
{Manchester}, R.~N., {et~al.} 2001, \mnras, 328, 17

\bibitem[{{Margalit} {et~al.}(2020){Margalit}, {Beniamini}, {Sridhar}, \&
  {Metzger}}]{mbs+20}
{Margalit}, B., {Beniamini}, P., {Sridhar}, N., \& {Metzger}, B.~D. 2020,
  \apjl, 899, L27

\bibitem[{{McEwen} {et~al.}(2024){McEwen}, {Lynch}, {Kaplan}, {Bolda},
  {Sengar}, {Fonseca}, {Agoudemos}, {Boyles}, {Chatterjee}, {Cohen},
  {Crawford}, {DeCesar}, {Ehlke}, {Fernandez}, {Ferrara}, {Fiore}, {Gilhaus},
  {Gleiter}, {Hessels}, {Holman}, {Joy}, {Kaspi}, {Kondratiev}, {Leon},
  {Levin}, {Lorenz}, {Lorimer}, {Madison}, {McLaughlin}, {Meyers}, {Parent},
  {Patron}, {Ransom}, {Ray}, {Roberts}, {Roch}, {Siemens}, {Stearns},
  {Swiggum}, {Stairs}, {Stovall}, {Tan}, {Valentine}, \& {van
  Leeuwen}}]{mlk+24}
{McEwen}, A.~E., {et~al.} 2024, \apj, 969, 118

\bibitem[{McKay {et~al.}(1979)McKay, Conover, \& Becker}]{LHS_paper}
McKay, M.~D., Conover, W.~J., \& Becker, R.~L. 1979, Technometrics, 21, 239

\bibitem[{{Melatos}(2000)}]{m00}
{Melatos}, A. 2000, \mnras, 313, 217

\bibitem[{{Metzger} {et~al.}(2018){Metzger}, {Beniamini}, \&
  {Giannios}}]{mbg18}
{Metzger}, B.~D., {Beniamini}, P., \& {Giannios}, D. 2018, \apj, 857, 95

\bibitem[{{Metzger} {et~al.}(2011){Metzger}, {Giannios}, {Thompson},
  {Bucciantini}, \& {Quataert}}]{2011MNRAS.413.2031M}
{Metzger}, B.~D., {Giannios}, D., {Thompson}, T.~A., {Bucciantini}, N., \&
  {Quataert}, E. 2011, \mnras, 413, 2031

\bibitem[{{Metzger} {et~al.}(2015){Metzger}, {Margalit}, {Kasen}, \&
  {Quataert}}]{mmk+15}
{Metzger}, B.~D., {Margalit}, B., {Kasen}, D., \& {Quataert}, E. 2015, \mnras,
  454, 3311

\bibitem[{Mishra-Sharma \& Cranmer(2022)}]{msc+22}
Mishra-Sharma, S., \& Cranmer, K. 2022, Phys. Rev. D, 105, 063017

\bibitem[{{Miyamoto} \& {Nagai}(1975)}]{mn+75}
{Miyamoto}, M., \& {Nagai}, R. 1975, \pasj, 27, 533

\bibitem[{{Navarro} {et~al.}(1997){Navarro}, {Frenk}, \& {White}}]{nfw+97}
{Navarro}, J.~F., {Frenk}, C.~S., \& {White}, S. D.~M. 1997, \apj, 490, 493

\bibitem[{{Noutsos} {et~al.}(2013){Noutsos}, {Schnitzeler}, {Keane}, {Kramer},
  \& {Johnston}}]{nsk+13}
{Noutsos}, A., {Schnitzeler}, D.~H.~F.~M., {Keane}, E.~F., {Kramer}, M., \&
  {Johnston}, S. 2013, \mnras, 430, 2281

\bibitem[{{Olausen} \& {Kaspi}(2014)}]{ok14}
{Olausen}, S.~A., \& {Kaspi}, V.~M. 2014, \apjs, 212, 6

\bibitem[{Oliphant(2006)}]{numpy}
Oliphant, T.~E. 2006, A guide to NumPy, Vol.~1 (Trelgol Publishing USA)

\bibitem[{{Omelyan} {et~al.}(2002){Omelyan}, {Mryglod}, \& {Folk}}]{omf+02}
{Omelyan}, I.~P., {Mryglod}, I.~M., \& {Folk}, R. 2002, \pre, 66, 026701

\bibitem[{{Palmer} {et~al.}(2005){Palmer}, {Barthelmy}, {Gehrels}, {Kippen},
  {Cayton}, {Kouveliotou}, {Eichler}, {Wijers}, {Woods}, {Granot}, {Lyubarsky},
  {Ramirez-Ruiz}, {Barbier}, {Chester}, {Cummings}, {Fenimore}, {Finger},
  {Gaensler}, {Hullinger}, {Krimm}, {Markwardt}, {Nousek}, {Parsons}, {Patel},
  {Sakamoto}, {Sato}, {Suzuki}, \& {Tueller}}]{palmer}
{Palmer}, D.~M., {et~al.} 2005, \nat, 434, 1107

\bibitem[{{Papamakarios} \& {Murray}(2016)}]{pm+16}
{Papamakarios}, G., \& {Murray}, I. 2016, arXiv e-prints, arXiv:1605.06376

\bibitem[{{Papamakarios} {et~al.}(2017){Papamakarios}, {Pavlakou}, \&
  {Murray}}]{ppm17}
{Papamakarios}, G., {Pavlakou}, T., \& {Murray}, I. 2017, arXiv e-prints,
  arXiv:1705.07057

\bibitem[{{Papamakarios} {et~al.}(2018){Papamakarios}, {Sterratt}, \&
  {Murray}}]{psm+18}
{Papamakarios}, G., {Sterratt}, D.~C., \& {Murray}, I. 2018, arXiv e-prints,
  arXiv:1805.07226

\bibitem[{{P{\'e}tri}(2012)}]{p+12}
{P{\'e}tri}, J. 2012, \mnras, 424, 605

\bibitem[{{Philippov} {et~al.}(2014){Philippov}, {Tchekhovskoy}, \&
  {Li}}]{ptl+14}
{Philippov}, A., {Tchekhovskoy}, A., \& {Li}, J.~G. 2014, \mnras, 441, 1879

\bibitem[{{Popov} {et~al.}(2010){Popov}, {Pons}, {Miralles}, {Boldin}, \&
  {Posselt}}]{ppm10}
{Popov}, S.~B., {Pons}, J.~A., {Miralles}, J.~A., {Boldin}, P.~A., \&
  {Posselt}, B. 2010, \mnras, 401, 2675

\bibitem[{{Popov} \& {Postnov}(2010)}]{popov_2010}
{Popov}, S.~B., \& {Postnov}, K.~A. 2010, in Evolution of Cosmic Objects
  through their Physical Activity, ed. H.~A. {Harutyunian}, A.~M. {Mickaelian},
  \& Y.~{Terzian}, 129--132

\bibitem[{{Price} {et~al.}(2021){Price}, {Flynn}, \& {Deller}}]{pygedm}
{Price}, D.~C., {Flynn}, C., \& {Deller}, A. 2021, \pasa, 38, e038

\bibitem[{{Price-Whelan} {et~al.}(2018){Price-Whelan}, {Sip{\H{o}}cz},
  {G{\"u}nther}, {Lim}, {Crawford}, {Conseil}, {Shupe}, {Craig}, {Dencheva},
  {Ginsburg}, {VanderPlas}, {Bradley}, {P{\'e}rez-Su{\'a}rez}, {de Val-Borro},
  {Paper Contributors}, {Aldcroft}, {Cruz}, {Robitaille}, {Tollerud},
  {Coordination Committee}, {Ardelean}, {Babej}, {Bach}, {Bachetti}, {Bakanov},
  {Bamford}, {Barentsen}, {Barmby}, {Baumbach}, {Berry}, {Biscani}, {Boquien},
  {Bostroem}, {Bouma}, {Brammer}, {Bray}, {Breytenbach}, {Buddelmeijer},
  {Burke}, {Calderone}, {Cano Rodr{\'\i}guez}, {Cara}, {Cardoso}, {Cheedella},
  {Copin}, {Corrales}, {Crichton}, {D{\textquoteright}Avella}, {Deil},
  {Depagne}, {Dietrich}, {Donath}, {Droettboom}, {Earl}, {Erben}, {Fabbro},
  {Ferreira}, {Finethy}, {Fox}, {Garrison}, {Gibbons}, {Goldstein}, {Gommers},
  {Greco}, {Greenfield}, {Groener}, {Grollier}, {Hagen}, {Hirst}, {Homeier},
  {Horton}, {Hosseinzadeh}, {Hu}, {Hunkeler}, {Ivezi{\'c}}, {Jain}, {Jenness},
  {Kanarek}, {Kendrew}, {Kern}, {Kerzendorf}, {Khvalko}, {King}, {Kirkby},
  {Kulkarni}, {Kumar}, {Lee}, {Lenz}, {Littlefair}, {Ma}, {Macleod},
  {Mastropietro}, {McCully}, {Montagnac}, {Morris}, {Mueller}, {Mumford},
  {Muna}, {Murphy}, {Nelson}, {Nguyen}, {Ninan}, {N{\"o}the}, {Ogaz}, {Oh},
  {Parejko}, {Parley}, {Pascual}, {Patil}, {Patil}, {Plunkett}, {Prochaska},
  {Rastogi}, {Reddy Janga}, {Sabater}, {Sakurikar}, {Seifert}, {Sherbert},
  {Sherwood-Taylor}, {Shih}, {Sick}, {Silbiger}, {Singanamalla}, {Singer},
  {Sladen}, {Sooley}, {Sornarajah}, {Streicher}, {Teuben}, {Thomas},
  {Tremblay}, {Turner}, {Terr{\'o}n}, {van Kerkwijk}, {de la Vega}, {Watkins},
  {Weaver}, {Whitmore}, {Woillez}, {Zabalza}, \& {Contributors}}]{astropy:2018}
{Price-Whelan}, A.~M., {et~al.} 2018, \aj, 156, 123

\bibitem[{{Rankin}(2007)}]{r+07}
{Rankin}, J.~M. 2007, \apj, 664, 443

\bibitem[{{Rea} {et~al.}(2015){Rea}, {Gull{\'o}n}, {Pons}, {Perna}, {Dainotti},
  {Miralles}, \& {Torres}}]{rgp+15}
{Rea}, N., {Gull{\'o}n}, M., {Pons}, J.~A., {Perna}, R., {Dainotti}, M.~G.,
  {Miralles}, J.~A., \& {Torres}, D.~F. 2015, \apj, 813, 92

\bibitem[{{Rea} {et~al.}(2008){Rea}, {Zane}, {Turolla}, {Lyutikov}, \&
  {G{\"o}tz}}]{rzt+08}
{Rea}, N., {Zane}, S., {Turolla}, R., {Lyutikov}, M., \& {G{\"o}tz}, D. 2008,
  \apj, 686, 1245

\bibitem[{{Ricci} {et~al.}(2021){Ricci}, {Troja}, {Bruni}, {Matsumoto}, {Piro},
  {O'Connor}, {Piran}, {Navaieelavasani}, {Corsi}, {Giacomazzo}, \&
  {Wieringa}}]{rtb+21}
{Ricci}, R., {et~al.} 2021, \mnras, 500, 1708

\bibitem[{{Riley} {et~al.}(2019){Riley}, {Watts}, {Bogdanov}, {Ray}, {Ludlam},
  {Guillot}, {Arzoumanian}, {Baker}, {Bilous}, {Chakrabarty}, {Gendreau},
  {Harding}, {Ho}, {Lattimer}, {Morsink}, \& {Strohmayer}}]{rwb+19}
{Riley}, T.~E., {et~al.} 2019, \apjl, 887, L21

\bibitem[{{Ronchi} {et~al.}(2021){Ronchi}, {Graber}, {Garcia-Garcia}, {Rea}, \&
  {Pons}}]{rgg+21}
{Ronchi}, M., {Graber}, V., {Garcia-Garcia}, A., {Rea}, N., \& {Pons}, J.~A.
  2021, \apj, 916, 100

\bibitem[{{Rozwadowska} {et~al.}(2021){Rozwadowska}, {Vissani}, \&
  {Cappellaro}}]{rvc21}
{Rozwadowska}, K., {Vissani}, F., \& {Cappellaro}, E. 2021, \na, 83, 101498

\bibitem[{{Safi-Harb} {et~al.}(2023){Safi-Harb}, {Burdge}, {Bodaghee}, {An},
  {Guest}, {Hare}, {Hebbar}, {Ho}, {Kargaltsev}, {Kirmizibayrak}, {Klingler},
  {Nynka}, {Reynolds}, {Sasaki}, {Sridhar}, {Vasilopoulos}, {Woods}, {Yang},
  {Heinke}, {Kong}, {Li}, {MacMaster}, {Mallick}, {Treyturik}, {Tsuji},
  {Binder}, {Braun}, {Chang}, {Chatterjee}, {Ferrand}, {Holland-Ashford}, {Ng},
  {Plotkin}, {Romani}, \& {Zhang}}]{sbb+23}
{Safi-Harb}, S., {et~al.} 2023, arXiv e-prints, arXiv:2311.07673

\bibitem[{{Sautron} {et~al.}(2024){Sautron}, {Pétri}, {Mitra}, \&
  {Dirson}}]{spm+24}
{Sautron}, M., {Pétri}, J., {Mitra}, D., \& {Dirson}, L. 2024, A\&A, 691, A349

\bibitem[{{Skowron} {et~al.}(2019){Skowron}, {Skowron}, {Mr{\'o}z}, {Udalski},
  {Pietrukowicz}, {Soszy{\'n}ski}, {Szyma{\'n}ski}, {Poleski}, {Koz{\l}owski},
  {Ulaczyk}, {Rybicki}, \& {Iwanek}}]{ssm+19}
{Skowron}, D.~M., {et~al.} 2019, Science, 365, 478

\bibitem[{{Smirnov}(1948)}]{kstest}
{Smirnov}, N. 1948, The Annals of Mathematical Statistics, 19, 279

\bibitem[{{Spitkovsky}(2006)}]{s+06}
{Spitkovsky}, A. 2006, \apjl, 648, L51

\bibitem[{{Spruit} \& {Phinney}(1998)}]{sp98}
{Spruit}, H., \& {Phinney}, E.~S. 1998, \nat, 393, 139

\bibitem[{{Stovall} {et~al.}(2014){Stovall}, {Lynch}, {Ransom}, {Archibald},
  {Banaszak}, {Biwer}, {Boyles}, {Dartez}, {Day}, {Ford}, {Flanigan}, {Garcia},
  {Hessels}, {Hinojosa}, {Jenet}, {Kaplan}, {Karako-Argaman}, {Kaspi},
  {Kondratiev}, {Leake}, {Lorimer}, {Lunsford}, {Martinez}, {Mata},
  {McLaughlin}, {Roberts}, {Rohr}, {Siemens}, {Stairs}, {van Leeuwen},
  {Walker}, \& {Wells}}]{slr+14}
{Stovall}, K., {et~al.} 2014, \apj, 791, 67

\bibitem[{{Tademaru} \& {Harrison}(1975)}]{th75}
{Tademaru}, E., \& {Harrison}, E.~R. 1975, \nat, 254, 676

\bibitem[{{Tauris} \& {Manchester}(1998)}]{tm98}
{Tauris}, T.~M., \& {Manchester}, R.~N. 1998, \mnras, 298, 625

\bibitem[{Tejero-Cantero {et~al.}(2020)Tejero-Cantero, Boelts, Deistler,
  Lueckmann, Durkan, Gonçalves, Greenberg, \& Macke}]{sbi-soft}
Tejero-Cantero, A., Boelts, J., Deistler, M., Lueckmann, J.-M., Durkan, C.,
  Gonçalves, P.~J., Greenberg, D.~S., \& Macke, J.~H. 2020, Journal of Open
  Source Software, 5, 2505

\bibitem[{{Totani} \& {Tsuzuki}(2023)}]{Totani2023}
{Totani}, T., \& {Tsuzuki}, Y. 2023, \mnras, 526, 2795

\bibitem[{{Trigg} {et~al.}(2024){Trigg}, {Burns}, {Roberts}, {Negro},
  {Svinkin}, {Baring}, {Wadiasingh}, {Christensen}, {Andreoni}, {Briggs}, {Di
  Lalla}, {Frederiks}, {Lipunov}, {Omodei}, {Ridnaia}, {Veres}, \&
  {Lysenko}}]{tbr+24}
{Trigg}, A.~C., {et~al.} 2024, \aap, 687, A173

\bibitem[{{Usov}(1992)}]{u92}
{Usov}, V.~V. 1992, \nat, 357, 472

\bibitem[{{Vasist} {et~al.}(2023){Vasist}, {Rozet}, {Absil}, {Molli{\`e}re},
  {Nasedkin}, \& {Louppe}}]{vra+23}
{Vasist}, M., {Rozet}, F., {Absil}, O., {Molli{\`e}re}, P., {Nasedkin}, E., \&
  {Louppe}, G. 2023, \aap, 672, A147

\bibitem[{{Vigan{\`o}}(2013)}]{v+13}
{Vigan{\`o}}, D. 2013, PhD thesis, University of Alacant, Spain

\bibitem[{{Vigan{\`o}} {et~al.}(2013){Vigan{\`o}}, {Rea}, {Pons}, {Perna},
  {Aguilera}, \& {Miralles}}]{vrp+13}
{Vigan{\`o}}, D., {Rea}, N., {Pons}, J.~A., {Perna}, R., {Aguilera}, D.~N., \&
  {Miralles}, J.~A. 2013, \mnras, 434, 123

\bibitem[{{Wang} {et~al.}(2024){Wang}, {Rea}, {Bao}, {Kaplan}, {Lenc},
  {Wadiasingh}, {Hare}, {Zic}, {Anumarlapudi}, {Bera}, {Beniamini}, {Cooper},
  {Clarke}, {Deller}, {Dawson}, {Glowacki}, {Hurley-Walker}, {McSweeney},
  {Polisensky}, {Peters}, {Younes}, {Bannister}, {Caleb}, {Dage}, {James},
  {Kasliwal}, {Karambelkar}, {Lower}, {Mori}, {Ocker}, {P{\'e}rez-Torres},
  {Qiu}, {Rose}, {Shannon}, {Taub}, {Wang}, {Wang}, {Zhao}, {Bhat}, {Dobie},
  {Driessen}, {Murphy}, {Jaini}, {Deng}, {Jahns-Schindler}, {Lee}, {Pritchard},
  {Tuthill}, \& {Thyagarajan}}]{Wang2024}
{Wang}, Z., {et~al.} 2024, arXiv e-prints, arXiv:2411.16606

\bibitem[{{Webb} {et~al.}(2020){Webb}, {Coriat}, {Traulsen}, {Ballet}, {Motch},
  {Carrera}, {Koliopanos}, {Authier}, {de la Calle}, {Ceballos}, {Colomo},
  {Chuard}, {Freyberg}, {Garcia}, {Kolehmainen}, {Lamer}, {Lin}, {Maggi},
  {Michel}, {Page}, {Page}, {Perea-Calderon}, {Pineau}, {Rodriguez}, {Rosen},
  {Santos Lleo}, {Saxton}, {Schwope}, {Tom{\'a}s}, {Watson}, \&
  {Zakardjian}}]{wct+20}
{Webb}, N.~A., {et~al.} 2020, \aap, 641, A136

\bibitem[{{Weltevrede} \& {Johnston}(2008)}]{wj08}
{Weltevrede}, P., \& {Johnston}, S. 2008, \mnras, 387, 1755

\bibitem[{{Wu} {et~al.}(2024){Wu}, {Damoulakis}, {Beniamini}, \&
  {Giannios}}]{wdb+24}
{Wu}, Z.-F., {Damoulakis}, M., {Beniamini}, P., \& {Giannios}, D. 2024, arXiv
  e-prints, arXiv:2411.12850

\bibitem[{{Yao} {et~al.}(2017){Yao}, {Manchester}, \& {Wang}}]{ymw+17}
{Yao}, J.~M., {Manchester}, R.~N., \& {Wang}, N. 2017, \apj, 835, 29

\bibitem[{{Younes} {et~al.}(2023){Younes}, {Baring}, {Harding}, {Enoto},
  {Wadiasingh}, {Pearlman}, {Ho}, {Guillot}, {Arzoumanian}, {Borghese},
  {Gendreau}, {G{\"o}{\v{g}}{\"u}{\c{s}}}, {G{\"u}ver}, {van der Horst}, {Hu},
  {Jaisawal}, {Kouveliotou}, {Lin}, \& {Majid}}]{ybh+23}
{Younes}, G., {et~al.} 2023, Nature Astronomy, 7, 339

\bibitem[{{Yusifov} \& {K{\"u}{\c{c}}{\"u}k}(2004)}]{yk+04}
{Yusifov}, I., \& {K{\"u}{\c{c}}{\"u}k}, I. 2004, \aap, 422, 545

\bibitem[{{Zane} {et~al.}(2009){Zane}, {Rea}, {Turolla}, \& {Nobili}}]{zrt+09}
{Zane}, S., {Rea}, N., {Turolla}, R., \& {Nobili}, L. 2009, \mnras, 398, 1403

\bibitem[{{Zhang} \& {Harding}(2000)}]{zh00}
{Zhang}, B., \& {Harding}, A.~K. 2000, \apjl, 535, L51

\bibitem[{{Zhou} {et~al.}(2014){Zhou}, {Chen}, {Li}, {Safi-Harb}, {Mendez},
  {Terada}, {Sun}, \& {Ge}}]{zcl+14}
{Zhou}, P., {Chen}, Y., {Li}, X.-D., {Safi-Harb}, S., {Mendez}, M., {Terada},
  Y., {Sun}, W., \& {Ge}, M.-Y. 2014, \apjl, 781, L16

\end{thebibliography}
